\documentclass[twocolumn]{aastex62}

\usepackage{amsmath}

\newcommand{\bmM}{{\boldsymbol{\mathcal{M}}}}
\newcommand{\mL}{{\mathcal{L}}}
\newcommand{\Var}{{\text{Var}}}
\newcommand{\Tr}{{\text{Tr}}}
\newcommand{\Real}{{\ \text{Re}}}
\newcommand{\Imag}{{\ \text{Im}}}
\newcommand{\mM}{{\mathcal{M}}}
\newcommand{\mN}{{\mathcal{N}}}
\newcommand{\bA}{{\boldsymbol{A}}}
\newcommand{\bB}{{\boldsymbol{B}}}
\newcommand{\bC}{{\boldsymbol{C}}}
\newcommand{\bD}{{\boldsymbol{D}}}
\newcommand{\bE}{{\boldsymbol{E}}}

\newcommand{\bF}{{\boldsymbol{F}}}
\newcommand{\bG}{{\boldsymbol{G}}}
\newcommand{\bI}{{\boldsymbol{I}}}

\newcommand{\bS}{{\boldsymbol{S}}}

\newcommand{\bq}{{\boldsymbol{q}}}

\newcommand{\bov}{{\boldsymbol{v}}}
\newcommand{\bn}{{\boldsymbol{n}}}
\newcommand{\bx}{{\boldsymbol{x}}}
\newcommand{\by}{{\boldsymbol{y}}}
\newcommand{\bJ}{{\boldsymbol{J}}}
\newcommand{\bgamma}{{\boldsymbol{\gamma}}}
\newcommand{\bxi}{{\boldsymbol{\xi}}}

\newcommand{\bDelta}{{\boldsymbol{\Delta}}}
\newcommand{\bPsi}{{\boldsymbol{\Psi}}}
\newcommand{\bsigma}{{\boldsymbol{\sigma}}}
\newcommand{\bSigma}{{\boldsymbol{\Sigma}}}
\newcommand{\bEs}{\bE_s}
\newcommand{\bEa}{\bE_a}
\newcommand{\bEatot}{\bE_{a,\text{tot}}}

\newcommand{\bSigmas}{\bSigma_s}
\newcommand{\bSigmaa}{\bSigma_a}
\newcommand{\bSigmaatot}{\bSigma_{a,\text{tot}}}

\newcommand{\Paave}{{P_{a,\text{ave}}}}

\newcommand{\Ex}{E_x}
\newcommand{\Ey}{E_y}
\newcommand{\JthetaX}{{J_{\theta x}}}
\newcommand{\JphiX}{{J_{\phi x}}}
\newcommand{\JthetaY}{{J_{\theta y}}}
\newcommand{\JphiY}{{J_{\theta y}}}
\newcommand{\Etheta}{{E_\theta}}
\newcommand{\Ephi}{{E_\phi}}
\newcommand{\brhat}{{\boldsymbol{\hat{r}}}}
\newcommand{\bxhat}{{\boldsymbol{\hat{x}}}}
\newcommand{\byhat}{{\boldsymbol{\hat{y}}}}
\newcommand{\bthetahat}{{\boldsymbol{\hat{\theta}}}}
\newcommand{\bphihat}{{\boldsymbol{\hat{\phi}}}}
\newcommand{\bzero}{{\boldsymbol{0}}}

\shorttitle{Global 21-cm signal extraction III: drift-scan and polarization}
\shortauthors{Tauscher et al.}

\begin{document}

\title{Global 21-cm signal extraction from foreground and instrumental effects III:

Utilizing drift-scan time dependence and full Stokes measurements}

\correspondingauthor{Keith Tauscher}

\author{Keith Tauscher}
\affiliation{Center for Astrophysics and Space Astronomy, Department of Astrophysical and Planetary Science, University of Colorado, Boulder, CO 80309, USA}
\affiliation{Department of Physics, University of Colorado, Boulder, CO 80309, USA}

\author{David Rapetti}
\affiliation{NASA Ames Research Center, Moffett Field, CA 94035, USA}
\affiliation{Universities Space Research Association, Mountain View, CA, 94043, USA}
\affiliation{Center for Astrophysics and Space Astronomy, Department of Astrophysical and Planetary Science, University of Colorado, Boulder, CO 80309, USA}

%\author{Jordan Mirocha}
%\affiliation{Department of Physics and Astronomy, University of California at Los Angeles, Los Angeles, CA 90095, USA}
%\affiliation{Department of Physics and McGill Space Institute, McGill University, Montreal, QC, Canada H3A 2T}

\author{Jack~O.~Burns}
\affiliation{Center for Astrophysics and Space Astronomy, Department of Astrophysical and Planetary Science, University of Colorado, Boulder, CO 80309, USA}

%\author{Eric Switzer}
%\affiliation{NASA Goddard Space Flight Center, Greenbelt, MD 20771, USA}

\email{Keith.Tauscher@colorado.edu}

\begin{abstract}
When using valid foreground and signal models, the uncertainties on extracted signals in global 21-cm signal experiments depend principally on the overlap between signal and foreground models. In this paper, we investigate two strategies for decreasing this overlap:~(i) utilizing time dependence by fitting multiple drift-scan spectra simultaneously and (ii) measuring all four Stokes parameters instead of only the total power, Stokes I. Although measuring polarization requires different instruments than are used in most existing experiments, all existing experiments can utilize drift-scan measurements merely by averaging their data differently. In order to evaluate the increase in constraining power from using these two techniques, we define a method for connecting Root-Mean-Square (RMS) uncertainties to probabilistic confidence levels.~Employing simulations, we find that fitting only one total power spectrum leads to RMS uncertainties at the few K level, while fitting multiple time-binned, drift-scan spectra yields uncertainties at the $\lesssim 10$ mK level. This significant improvement only appears if the spectra are modeled with one set of basis vectors, instead of using multiple sets of basis vectors that independently model each spectrum. Assuming that they are simulated accurately, measuring all four Stokes parameters also leads to lower uncertainties. These two strategies can be employed simultaneously and fitting multiple time bins of all four Stokes parameters yields the best precision measurements of the 21-cm signal, approaching the noise level in the data.
\end{abstract}

\keywords{cosmology: dark ages, reionization, first stars --- cosmology: observations}

\section{Introduction}
\label{sec:introduction}

The hyperfine, spin-flip transition of neutral hydrogen produces radiation of 1420 MHz of frequency in the rest frame, corresponding to a wavelength of 21 cm \citep{Hellwig:70}. Although this transition is highly forbidden, with a mean lifetime of around 11 million years \citep[][Section 7.8]{Condon:16}, its emission and absorption are visible from the vast amount of neutral gas in the early universe, redshifted to low frequencies of 10-200 MHz by cosmic expansion \citep{Pritchard:12}. It is the only existing direct probe of the neutral hydrogen in the Dark Ages and Cosmic Dawn of the early Universe and it could be a powerful tool in the study of the Epoch of Reionization, when the hydrogen in the Universe was ionized by light from compact sources like stars and black holes \citep{Furlanetto:06}. Two aspects of this 21-cm signal are currently under study: the power spectrum, where angular variations in the gas evolution manifest \citep{Morales:10}, and the sky-averaged (global) monopole component, which tracks the average properties of the gas across the Universe as a function of cosmic time \citep{Pritchard:10}. This paper concerns the latter.

The most difficult analysis task in measuring the global 21-cm signal is separating it from foreground emission from our galaxy that is $\sim 10^{4-6}$ times larger than the signal, which is expected to have an amplitude of a few hundred mK. The foreground emission largely consists of synchrotron radiation, which follows a power law in frequency when the energy of the electrons emitting it follows a power law distribution \citep[][section 5.2]{Condon:16}; so it is expected to be very spectrally smooth. However, there are large anisotropies in galactic emission both in magnitude and spectral index, which are averaged together by wide antenna beams that also change in frequency. Due to the corruption caused by this beam averaging, there is no obvious analytical model to use to fit the beam-weighted foreground spectrum, although many have used polynomial-based models \citep{SathyanarayanaRao:17,Monsalve:17,Bowman:18}.

In Paper I of this series \citep{Tauscher:18}, we laid out a procedure for extracting the global signal from foregrounds without assuming a particular foreground model, but instead by simulating the foregrounds many times, with the parameters of these simulations varying between limits corresponding to realistic uncertainties. Using these simulations as a training set of foregrounds, the pipeline performs Singular Value Decomposition (SVD) to extract orthogonal basis vectors with which to fit the foreground. After performing the same process with the (much wider) training set of global signals, we fit the spectral data simultaneously with both SVD models and use the signal basis and the corresponding fit coefficients to construct confidence intervals on the 21-cm signal. The uncertainties on these intervals depend on the noise level of the data and the overlap between foreground and signal basis vectors.

In this paper, we use our pipeline to show that the overlap between foreground and 21-cm signal can be mitigated by utilizing time dependent drift-scan measurements and observations of the four Stokes parameters describing polarization. The ability to use these extra pieces of data efficiently in constraining the signal is unique to our pipeline.~While one can perform inference on drift-scan measurements using a polynomial-based method, the connection between the foregrounds of the spectra (i.e., the fact that they come from the same beam and sky offset by some angle) cannot be fully accounted for. There is also no clear way to extend polynomial methods to Stokes parameters while utilizing the connection between them to help constrain the signal.

Our method of using SVD, which, for the present purposes, is equivalent to the diagonalization of a covariance matrix, to produce basis vectors is similar to past work performed in 21-cm cosmology. For example, \cite{Switzer:14} estimated eigenmodes of the foreground frequency covariance matrix from data taken at different times (or, equivalently, pointing directions). In our work, the modes are generated from a priori training sets based on previously and independently observed foreground spectra and simulated and/or measured beams instead of the sky-averaged radio spectra themselves. While utilizing the data to find modes is tempting because it relies less on a priori information, it is complicated by the fact that the foreground is never observed without the signal included.

Another method similar to the one discussed here is presented in \cite{Vedantham:14}, henceforth denoted as V14, where the authors simulate foreground spectra at different time snapshots, stack the resulting spectra into a matrix, and perform SVD to retrieve eigenmodes. The key difference between this and our method is that V14 only derives modes as a function of frequency by performing SVD on one simulation of spectra from a series of times, whereas our technique utilizes multiple individual simulations, each of which contains spectra for a series of times, to produce modes that differ both as a function of frequency and as a function of time. The correlation of the spectra from time to time is key to include in the model of the beam-weighted foreground data, as will be shown in Section~\ref{sec:results}, because even if the frequency modes of V14 can fit the beam-weighted foreground well, using them independently in each time-binned spectrum leads to a large overlap between foreground and signal models, producing large uncertainties. The extremely important role that utilizing spectrum-to-spectrum correlations caused by angular variations in the foreground play in precise measurements of the global 21-cm signal was seen clearly by \cite{Liu:13}, which also, similarly to the pipeline first presented in Paper I, provided a generalizable method of producing a generic linear basis for the beam-weighted foreground across different angles through diagonalization of a covariance matrix.

In Paper II \citep{Rapetti:19}, we presented our pipeline's strategy to translate from spectral constraints to nonlinear signal parameter constraints using a Markov Chain Monte Carlo (MCMC) algorithm, while analytically marginalizing over the same SVD-derived modes for the foreground as used in forming the spectral constraints at each step. This allows us to efficiently explore the MCMC parameter space of the nonlinear signal, fully accounting for complex foreground models from many correlated spectra. The latter is critical to extract the signal at the level required by standard 21-cm models, as we demonstrate in this paper, the third of the series.

In Section~\ref{sec:pipeline-review}, we review the pipeline, with a particular focus on how the overlap between signal and foreground generates uncertainties in the signal extraction. In Section~\ref{sec:driftscan}, we present how we simulate training sets using drift-scan measurements and how they help reduce overlap between foreground and signal. In Section~\ref{sec:polarization}, we do the same for measurements of the Stokes parameters by pairs of dipoles. In Section~\ref{sec:simulations}, we describe the simulation setup with which we test the benefits of including drift-scan and polarization measurements. In Section~\ref{sec:results}, we connect RMS uncertainties to confidence levels and compare the uncertainties with and without polarization and drift-scan measurements. We conclude in Section~\ref{sec:conclusions}.

\section{Pipeline review}
\label{sec:pipeline-review}

\subsection{Formalism}

The basis of our pipeline is the formation of the data vector, $\by$, which contains a large number of individual spectra concatenated,
\begin{equation}
  \by = \by_{\text{fg}} + \bPsi_{21}\by_{21} + \bn
\end{equation}
where $\by_{\text{fg}}$ and $\by_{21}$ are the true foreground and signal vectors, respectively, $\bn$ is a random Gaussian noise vector with covariance $\bC$, and $\bPsi_{21}$ is the so-called ``signal expansion matrix,'' explained further below. $\by$, $\by_{\text{fg}}$, and $\bn$ are vectors of length $n_cn_\nu$, where $n_c$ is the number of concatenated spectra in the data and $n_\nu$ is the number of frequencies in each spectrum. Since the signal is a single spectrum (i.e.~a vector of length $n_\nu$), it must be expanded into the full, length-$n_cn_\nu$ space of $\by$.~Expanding the signal into the dimensions of the full data vector while encoding information on how the data were obtained is the purpose of the signal expansion matrix $\bPsi_{21}$. Because $\bPsi_{21}\by_{21}$ must be a length-$n_cn_\nu$ vector and $\by_{21}$ is a length-$n_\nu$ vector, $\bPsi_{21}$ is an $n_cn_\nu\times n_\nu$ matrix. Examples of the signal expansion matrix in specific circumstances are provided in Sections~\ref{sec:expmatrix_drift},~\ref{sec:expmatrix_pol},~and~\ref{sec:signal-training-set}.\footnote{In some applications, it is useful to define expansion matrices for more components than just the signal. For some examples, see Paper I. In this case, since the beam-weighted foreground training set is made of many sets of spectra covering the whole data space, there is no need for a foreground expansion matrix.}

We model the data using weighted combinations of basis vectors contained in matrices denoted $\bF_{\text{fg}}$ and $\bF_{21}$, composed of the singular vectors of the foreground and signal training sets, respectively.~These matrices are found via SVD and are normalized such that $\bF_{\text{fg}}^T\bC^{-1}\bF_{\text{fg}}=\bI$ and $\bF_{21}^T\bPsi_{21}^T\bC^{-1}\bPsi_{21}\bF_{21}=\bI$, where $\bI$ is the identity matrix. The model of the data is
\begin{equation}
  \bmM(\bx_{\text{fg}},\bx_{21}) = \bF_{\text{fg}}\bx_{\text{fg}} + \bPsi_{21}\bF_{21}\bx_{21},
\end{equation}
where $\bx_{\text{fg}}$ and $\bx_{21}$ are weighting coefficients for the foreground and signal basis vectors, respectively. This is the same as $\bmM=\bG\bx$ where $\bG=\begin{bmatrix}\bF_{\text{fg}} & \bPsi_{21}\bF_{21}\end{bmatrix}$ and $\bx^T=\begin{bmatrix} \bx_{\text{fg}}^T & \bx_{21}^T \end{bmatrix}$. The probability distribution of the parameters is then taken to be proportional to the likelihood, given by
\begin{equation}
    \mL(\bx)\propto\exp{\left\{-\frac{1}{2}(\by-\bG\bx)^T\bC^{-1}(\bG\bx-\by)\right\}}.
\end{equation}
This implies that $\bx$ is normally distributed with mean $\bxi$ and covariance $\bS$ where
\begin{equation}
  \bS = (\bG^T\bC^{-1}\bG)^{-1} \ \ \text{ and } \ \ \bxi=\bS\bG^T\bC^{-1}\by.
\end{equation}
We then create signal confidence intervals centered on $\bgamma_{21}$ with a channel covariance $\bDelta_{21}$ given by
\begin{subequations}
\begin{align}
  \bgamma_{21} &= \bF_{21} \bxi_{21}~, \label{eq:channel-mean} \\
  \bDelta_{21} &= \bF_{21} \bS_{21} \bF_{21}^T, \label{eq:channel-covariance}
\end{align} \label{eq:channel-mean-and-covariance}
\end{subequations}
where $\bxi_{21}$ and $\bS_{21}$ are the parts of $\bxi$ and $\bS$ corresponding to the signal parameters. The 1-sigma Root-Mean-Square (RMS) uncertainty on the signal can then be defined as
\begin{equation}
  \text{RMS}^{1\sigma}_{21} = \sqrt{\frac{\Tr(\bDelta_{21})}{n_\nu}}\,. \label{eq:channel-rms}
\end{equation}
This mathematical formalism is implemented in the \texttt{pylinex} Python code.\footnote{\cite{pylinex:20}, current version at \url{https://bitbucket.org/ktausch/pylinex}}

\subsection{Effect of overlap on uncertainties}

From the reconstruction described by Equations~\ref{eq:channel-mean-and-covariance}, we can define the normalized RMS error on the signal as $\text{NRMS}_{21}=\sqrt{\Tr(\bC^{-1/2}\bPsi_{21}\bDelta_{21}\bPsi_{21}^T\bC^{-1/2})/n_\nu}$, which is essentially the RMS of the ratio of the 1$\sigma$ uncertainty level to the 1$\sigma$ noise level, leading to a unitless summary quantity that is 1 if the 1$\sigma$ posterior uncertainty level is the same size as the 1$\sigma$ noise level. It is given by
\begin{subequations} \begin{align}
    \text{NRMS}_{21} &= \sqrt{\frac{\Tr(\bS_{21}\bF_{21}^T\bPsi_{21}^T\bC^{-1}\bPsi_{21}\bF_{21})}{n_\nu}}, \\
    &= \sqrt{\frac{\Tr(\bS_{21})}{n_\nu}}\;. \label{eq:normalized-channel-rms}
\end{align} \end{subequations}
Through block inversion, it is possible to compute that $\bS_{21}=(\bI-\bD^T\bD)^{-1}$ where $\bD=\bF_{\text{fg}}^T\bC^{-1}\bPsi_{21}\bF_{21}$ is the matrix of overlaps (dot products) between the foreground and signal basis vectors.  The trace of $\bS_{21}$ is therefore $\sum_{j=1}^{n_{21}}\frac{1}{1-\lambda_j}$ where $n_{21}$ is the number of signal vectors and $\lambda_j$ are the eigenvalues of $\bD^T\bD$.\footnote{This follows from the fact that $\Tr[(\bI-\bD^T\bD)^{-1}]=\Tr[\sum_{k=0}^{\infty}(\bD^T\bD)^k]=\sum_{k=0}^{\infty}\Tr[(\bD^T\bD)^k]=\sum_{k=0}^\infty\sum_{j=1}^{n_{21}}{\lambda_j}^k=\sum_{j=1}^{n_{21}}\sum_{k=0}^\infty{\lambda_j}^k=\sum_{j=1}^{n_{21}}\frac{1}{1-\lambda_j}$.} Thus,
\begin{equation}
    \text{NRMS}_{21}=\sqrt{\frac{1}{n_\nu}\sum_{j=1}^{n_{21}}\frac{1}{1-\lambda_j}}.
\end{equation}

If all foreground and signal basis vectors are orthogonal (i.e. $\bD=\bzero$), then the eigenvalues are all zero and $\text{NRMS}_{21}$ reaches its minimum value of $\sqrt{n_{21}/n_\nu}$. If, on the other hand, at least one foreground vector can be written as a combination of the signal vectors, or vice versa, then at least one of the eigenvalues of $\bD^T\bD$ is 1 and $\text{NRMS}_{21}$ diverges to $\infty$. In general $\text{NRMS}_{21}$ lies between these two extremes. While we utilize $\text{RMS}^{1\sigma}_{21}$ to report results in this paper, its normalized version $\text{NRMS}_{21}$ is useful to illustrate how the overlap between the signal and foreground vectors leads to greater uncertainty. The same effect is present when computing $\text{RMS}^{1\sigma}_{21}$ but cannot be shown as clearly analytically.

\begin{figure}[tb]
    \centering
    \includegraphics[width=0.233\textwidth]{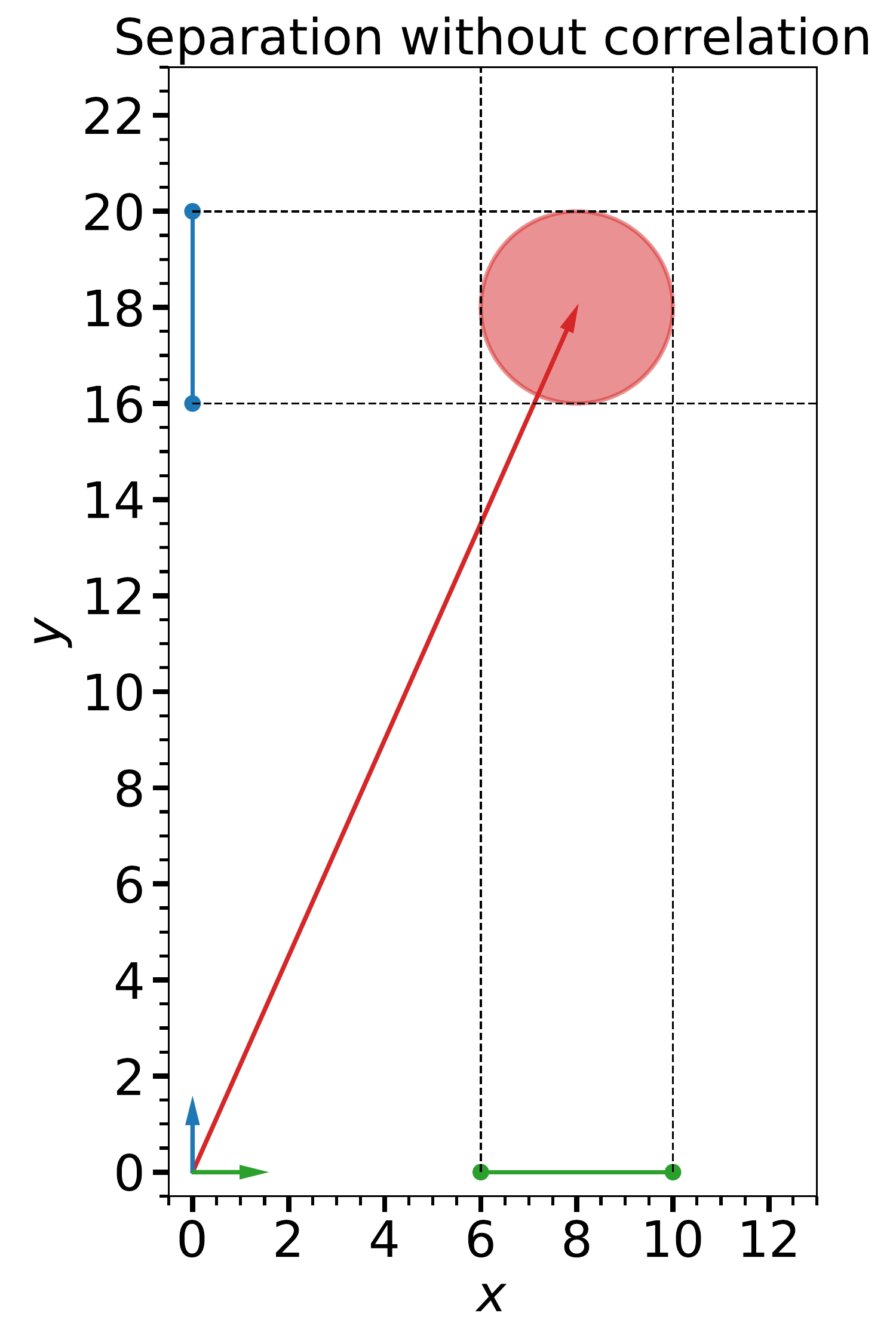}
    \includegraphics[width=0.233\textwidth]{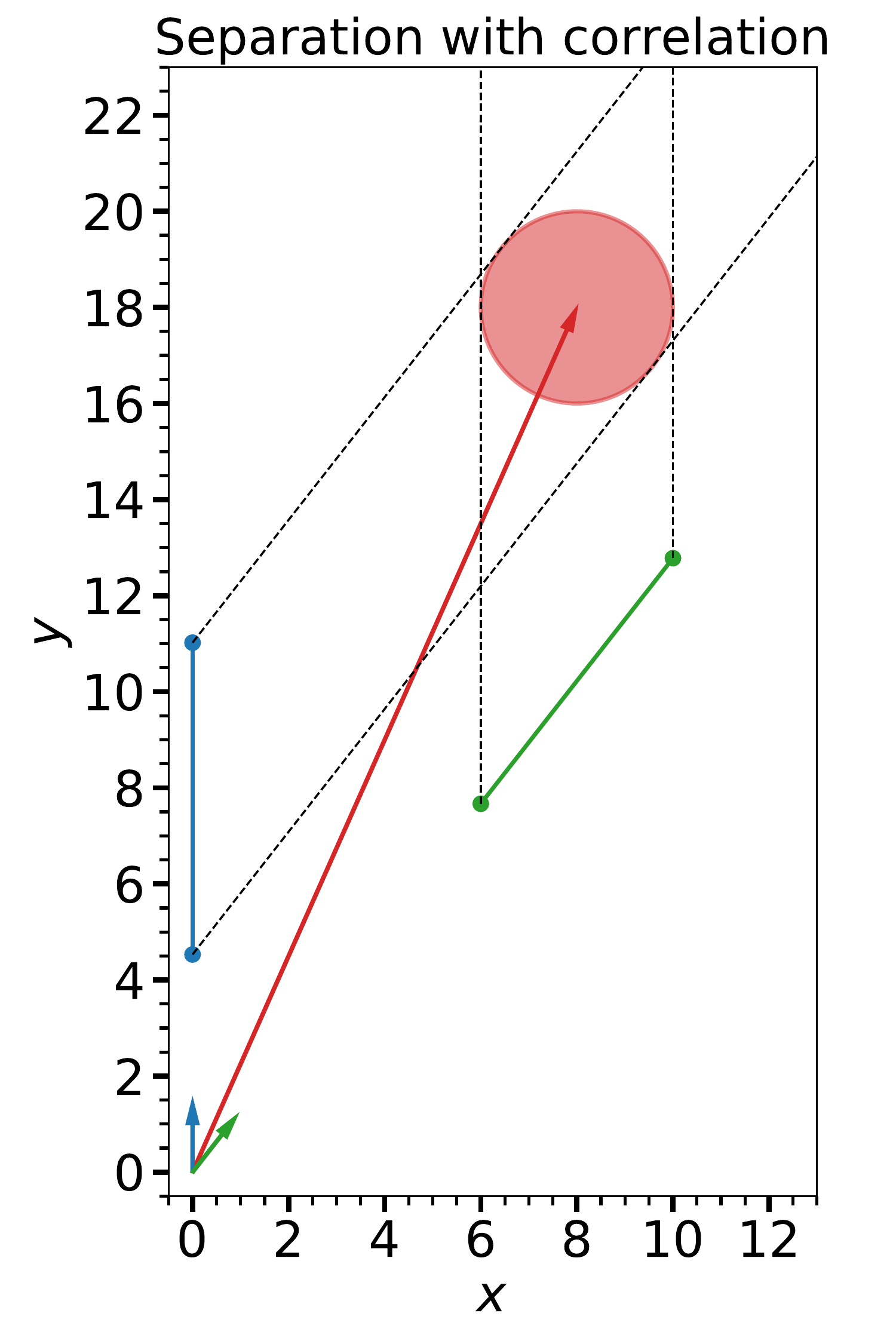}
    \caption{Simplified schematic representation of how the overlap between signal and foreground modes increases the uncertainties of both individually separated components with respect to the minimum level determined by the statistical noise. The red circle represents the $2\sigma$ noise uncertainty of the data (red vector). The blue and green vectors whose tails sit on the origin represent the signal and foreground basis vectors, respectively. The blue and green intervals demarcated by solid circles are the $2\sigma$ uncertainties on the signal and foreground. The signal (foreground) uncertainty is computed by projecting the noise ellipse parallel to the foreground (signal) basis vector onto the line defined by the signal (foreground) basis vector. The left panel shows that the minimum uncertainties for each of the components, defined by the noise level, is achieved by using orthogonal modes, while larger uncertainties are obtained when the overlap is large, as occurs in the right panel. In the left (right) panel, the dot product between the signal and foreground unit vectors, which is also the y-coordinate of the green unit vector, is 0 (0.7). In this simplified example, the 1D uncertainties are proportional to $|\csc{\alpha}|$ where $\alpha$ is the angle between the unit vectors.}
    \label{fig:vector_separation}
\end{figure}

Figure~\ref{fig:vector_separation} shows a schematic explanation of how noise in data interacts with the overlap between signal and foreground basis vectors, confusing the extraction of the signal. In this case, $\bPsi_{21}$ is the identity matrix because the foreground and signal exist in the same space. The standard deviations of the one dimensional confidence intervals on foreground and signal (lengths of blue and green line segments) are projections of the noise (red ellipse) onto the foreground and signal basis in a manner perpendicular to the other basis. In this simple case of two unit vectors, $\text{NRMS}_{21}\propto\csc{\alpha}$ where $\alpha$ is the angle between the unit vectors. So, as in the general case, as the basis vectors get closer to each other ($\alpha$ gets smaller), the uncertainties grow.

\section{Time dependence with drift-scan}
\label{sec:driftscan}

\subsection{Drift-scan formalism}

To simulate drift-scan measurements for training sets made to fit data from a ground-based experiment, we compute the boresight direction of a zenith-pointing antenna at a given latitude, longitude, and Local Sidereal Time (LST). Using this direction and the orientation of the antenna with respect to geographic north, we can define a foreground power map that is a function of sky position (given in terms of antenna-based spherical coordinate angles $\theta$ and $\phi$), frequency, $\nu$, and sidereal time, $t$, as $T(\theta,\phi,\nu,t)$. Any real observation will take place over a finite time period, say from $t_i$ to $t_f$. The effective foreground seen by the antenna is a smeared version of the foreground created by an integral of $T$, given by
\begin{equation}
  T_{\text{eff}}(\theta,\phi,\nu,t_i\rightarrow t_f) = \frac{1}{t_f-t_i}\int_{t_i}^{t_f}T(\theta,\phi,\nu,t)\ dt.
\end{equation}
In practice, we split the time interval into $n+1$ snapshots, so that the integral can be approximated by the following finite Riemann sum:
\begin{equation}
  T_{\text{eff}}(\theta,\phi,\nu,t_i\rightarrow t_f) = \frac{1}{n+1}\sum_{k=0}^nT\left(\theta,\phi,\nu,t_k\right)\,,
\end{equation}
where $t_k=t_i+\frac{k}{n}(t_f-t_i)$.

\subsection{Drift-scan expansion matrix}
\label{sec:expmatrix_drift}

While the foreground changes as a function of time, the global 21-cm signal exists equally in every spectrum when using a drift-scan measurement strategy. Therefore, if there are $n_{\text{drift}}$ measured spectra, the drift-scan expansion matrix for the 21-cm signal is
\begin{equation}
  \bPsi_{21,\text{drift}}^T = \underbrace{\begin{bmatrix} \bI & \bI & \ldots & \bI \end{bmatrix}}_{n_{\text{drift}}\ \bI\text{'s}}
\end{equation}
where $\bI$ is the identity matrix.~Because the signal does not change as the foreground changes, drift-scan measurements decrease the similarity between the foreground and signal models.

\section{Observation of Stokes parameters}
\label{sec:polarization}

Full Stokes measurements provide another excellent mechanism for reducing overlap between signal and foreground modes because foreground modes appear in all polarization modes while the global 21-cm signal appears only in Stokes I due to its lack of polarization and its isotropy. To include Stokes parameters in any analysis, however, one must first accurately simulate observations including them. This section presents a formalism for simulating full Stokes observations, splitting the results up into two terms, the induced polarization term that comes from projection of unpolarized radiation onto the antenna plane and the intrinsic polarization term which comes from polarized foreground sources. Simulations using this formalism will later be used to generate training sets with the purpose of computing modes that encode correlations between the different Stokes parameters and with which to fit the beam-weighted foreground.

In this section, we outline methods of performing simulations of beam-weighted foreground measurements based on the Jones matrix \citep[see][]{Jones:41}, which was first introduced to describe polarization measurements and coordinate transformations for optical systems, but has since been used in addition for other wavelengths, such as at CMB frequencies \citep[see, e.g.][]{ODea:07,Chuss:12}.

Data from radio antennas are caused by electric fields, $\bEs$, from the sky, which are written in terms of $\theta$ and $\phi$ components, $\Etheta$ and $\Ephi$, i.e. $\bEs=\Etheta\bthetahat+\Ephi\bphihat$ (note that there is no $\brhat$ component of the electric field because the radiation is traveling in the $-\brhat$ direction), where $\theta=0$ is the pointing direction of the antenna.~$\Etheta$ and $\Ephi$ can in general be complex and $\bEs$ be a complex random vector. The Stokes parameters of the sky radiation, $I_s$, $Q_s$, $U_s$, and $V_s$, which are the real power-unit quantities measuring polarization, are then given by
\begin{subequations} \begin{align}
  I_s &= \left\langle|\Etheta|^2 + |\Ephi|^2\right\rangle \\
  Q_s &= \left\langle|\Etheta|^2 - |\Ephi|^2\right\rangle \\
  U_s &= \left\langle 2\Real(E_{\theta}^\ast E_{\phi})\right\rangle \\
  V_s &= \left\langle 2\Imag(E_{\theta}^\ast E_{\phi})\right\rangle
\end{align} \end{subequations}
where $\langle\ldots\rangle$ denotes the expectation value. This can be written as $P_s=\langle\bE_s^\dagger\bsigma_P\bE_s\rangle$ where $\dagger$ represents the Hermitian transpose and
\begin{subequations} \label{eq:pauli-matrices} \begin{align}
  \bsigma_I = \begin{bmatrix} 1 & 0 \\ 0 & 1 \end{bmatrix},\ \ \ 
  \bsigma_Q = \begin{bmatrix} 1 & 0 \\ 0 & -1 \end{bmatrix}, \\ 
  \bsigma_U = \begin{bmatrix} 0 & 1 \\ 1 & 0 \end{bmatrix},\ \ \ 
  \bsigma_V = \begin{bmatrix} 0 & -i \\ i & 0 \end{bmatrix},
\end{align} \end{subequations}
are the Pauli matrices \citep{Fano:54}.~$\bEs$ is a function of both sky position and frequency, so $P_s$ is as well.

\subsection{Sky polarization}
\label{sec:sky-polarization}

Assuming that there is no coherent radiation coming from the sky, the expectation value of the electric field is zero, $\langle\bEs\rangle=0$. Since $\bEs$ is coming from many different electrons (in the case of synchrotron emission), and every phase is equally probable, $\bEs$ follows a circularly symmetric complex normal distribution with probability density
\begin{equation}
  f(\bEs) = \frac{\exp{\left(-\bEs^\dagger\bSigmas^{-1}\bEs\right)}}{\pi^2|\bSigmas|},
\end{equation}
where $\bSigmas=\langle\bEs\bEs^\dagger\rangle$ is the Hermitian covariance matrix. With this probability density, the expected values of the Stokes parameters are given by\footnote{See Appendix~\ref{app:Stokes-noise} for an example of how to compute expectation values with this probability density form.}
\begin{equation}
  P_s = \Tr(\bsigma_P\bSigmas). \label{eq:Stokes-as-trace}
\end{equation}

Since the distribution of $\bEs$ can represent any elliptical shape around the origin, it can be decomposed into the sum of two independent normally distributed vectors, one with a circular covariance matrix (i.e., proportional to the identity matrix) and another that exists only along a line, specified by a complex vector $\bov_s$, satisfying $\bov_s^\dagger\bov_s=1$.~This means that $\bSigmas$ can be written as $\bSigmas = \alpha_s\bI + \beta_s\bov_s\bov_s^\dagger$, where $\bI$ is the $2\times 2$ identity matrix and $\alpha_s$ and $\beta_s$ are non-negative, so $P_s = \alpha_s\Tr(\bsigma_P) + \beta_s\bov_s^\dagger\bsigma_P\bov_s$. To interpret $\alpha_s$ and $\beta_s$, we write the expression for the total intensity of the sky radiation, $I_s$, by plugging in $\bsigma_P=\bsigma_I=\bI$ and using $\bov_s^\dagger\bov_s=1$.~We find $I_s = 2\alpha_s+\beta_s$. Since $\alpha_s$ is the coefficient in front of the circular covariance matrix, it must involve only unpolarized radiation; so, we write $\alpha_s=[(1-p_s)/2]I_s$ where $0\le p_s\le 1$ is the polarization fraction of the sky radiation, leaving us with $\beta_s=p_sI_s$. This means that $\bSigmas=[(1-p_s)/2]I_s\bI+p_sI_s\bov_s\bov_s^\dagger$ and
\begin{equation}
    P_s = \left(\frac{1-p_s}{2}\right)I_s\ \Tr(\bsigma_P) + p_sI_s\bov_s^\dagger\bsigma_P\bov_s. \label{eq:sky-polarization}
\end{equation}
$I_s$ in these expressions can be taken from total power maps of the sky at a given frequency. $p_s$ and $\bov_s$ can be determined from $Q_s$, $U_s$, and $V_s$ using Equation~\ref{eq:sky-polarization} and noting that $\Tr(\bsigma_Q)=\Tr(\bsigma_U)=\Tr(\bsigma_V)=0$. If there is no circular polarization coming from the sky, $V_s=0$, implying that both components of $\bov_s$ have the same phase, meaning that, up to an arbitrary phase, it can be expressed through $\bov_s^\dagger=\begin{bmatrix} \cos{\psi_s} & \sin{\psi_s} \end{bmatrix}$. Plugging this expression into Equation~\ref{eq:sky-polarization}, $Q_s$ and $U_s$ can be written in the $V_s=0$ case as
\begin{equation}
  Q_s+iU_s = p_sI_se^{2i\psi_s}.
\end{equation}
Therefore, in this case,
\begin{equation}
  p_s = \left|\frac{Q_s+iU_s}{I_s}\right| \ \ \text{ and } \ \ \psi_s = \frac{1}{2}\ \text{Arg}\left(Q_s+iU_s\right).
\end{equation}

The random vector $\bEs$ can be written as the sum of two independent random vectors, $\bA_s$ with covariance $[(1-p_s)/2]I_s\bI$ and $\bB_s$ with covariance $p_sI_s\bov_s\bov_s^\dagger$. Both $\bA_s$ and $\bB_s$ contribute to $I_s$, but only $\bB_s$ contributes to $Q_s$ and $U_s$.

\subsection{Antenna polarization}
\label{sec:antenna-polarization}

The electric fields induced in the antenna can be written as $\bEa=\Ex\bxhat+\Ey\byhat$ where $\bxhat$ and $\byhat$ are the (generally orthogonal) antenna polarization directions. $\bEa$ is derived from $\bEs$ through a matrix known as the Jones matrix, $\bJ$.
\begin{equation}
  \bEa=\bJ\bEs, \label{eq:Jones-matrix-definition}
\end{equation}
or, equivalently,
\begin{equation}
  \begin{bmatrix} \Ex \\ \Ey \end{bmatrix} = \begin{bmatrix} \JthetaX & \JphiX \\ \JthetaY & \JphiY \end{bmatrix} \begin{bmatrix} \Etheta \\ \Ephi \end{bmatrix}.
\end{equation}
We can now solve for the Stokes parameters seen by the antennas by using the complex random vectors $\bA_s$ and $\bB_s$ in Equation~\ref{eq:Jones-matrix-definition}, i.e. $\bEa=\bJ(\bA_s+\bB_s)$. This implies
\begin{subequations} \begin{align}
  P_a &= \langle\bEa^\dagger\bsigma_P\bEa\rangle, \\
  %&= \langle (\bA_s+\bB_s)^\dagger\bJ^\dagger\bsigma_P\bJ(\bA_s+\bB_s) \rangle, \\
  &= \langle\bA_s^\dagger\bJ^\dagger\bsigma_P\bJ\bA_s\rangle + \langle\bB_s^\dagger\bJ^\dagger\bsigma_P\bJ\bB_s\rangle,
\end{align} \end{subequations}
where the last line follows because $\bA_s$ and $\bB_s$ are zero-mean and independent. Using the covariances of $\bA_s$ and $\bB_s$ derived in Section~\ref{sec:sky-polarization}, this can be written
\begin{multline}
  P_a = \left(\frac{1-p_s}{2}\right)I_s\ \Tr(\bJ^\dagger\bsigma_P\bJ) + p_sI_s\bov_s^\dagger\bJ^\dagger\bsigma_P\bJ\bov_s. \label{eq:antenna-polarization-two-terms}
\end{multline}

\begin{figure}[t!!]
  \centering
  \includegraphics[width=0.46\textwidth]{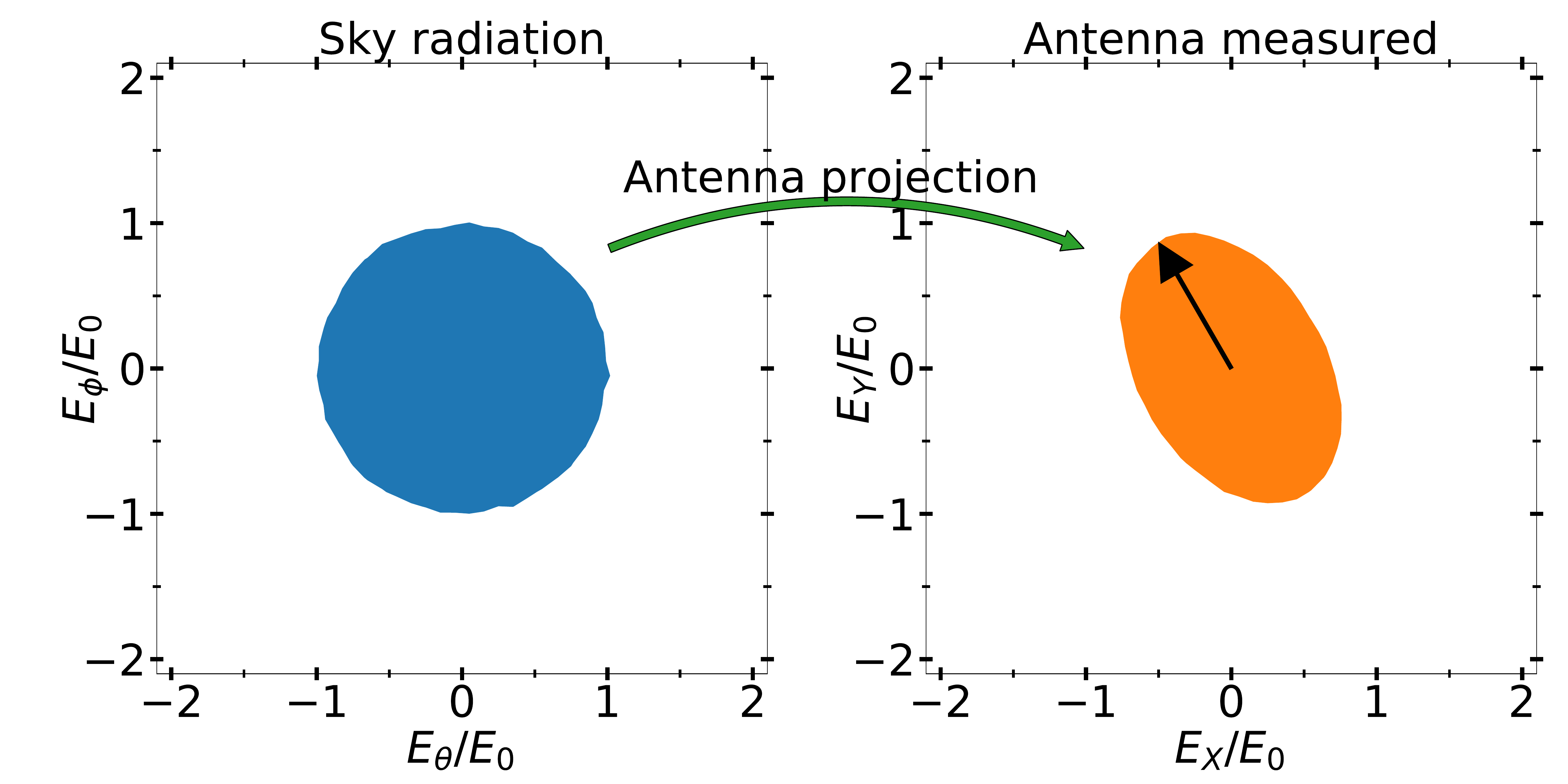}
  \caption{Illustration of the induced polarization effect for orthogonal ideal dipoles.~\textit{Left}:~Unpolarized sky radiation coming from $\theta=50^\circ$ off boresight and $\phi=30^\circ$ from the direction of the X-antenna, given in terms of $\theta$ and $\phi$ electric field components. The filled blue region represents the $1\sigma$ confidence interval with a covariance matrix of $(I_s/2)\bI$ where $I_s=2{E_0}^2$.~\textit{Right}: The projection of the electric fields from the left panel onto the X- and Y-antennas using the Jones matrix of orthogonal ideal dipoles, defined in Equation~\ref{eq:intuitive-Jones-matrix}. The filled orange region represents the $1\sigma$ confidence interval with a covariance matrix of $[(1-p_a)/2]I_a\bI+p_aI_a\bov_a\bov_a^\dagger$ where $I_a=(I_s/2)(2-\sin^2{\theta})$ (given by Equation~\ref{eq:total-power-all-terms} with $p_s=0)$, $p_a=\sin^2{\theta}/(2-\sin^2{\theta})$, and $\bov_a$ is a unit vector in the direction of the black arrow.
  } \label{fig:induced-polarization}
\end{figure}

As opposed to the sky polarization case, in general, both of these terms contribute to $I_a$, $Q_a$, $U_a$, and $V_a$. The first term encodes Stokes parameters induced from the unpolarized radiation from the sky while the second term encodes the effect of polarization intrinsic to the sky, so we term them induced and intrinsic polarization, respectively. In Appendix~\ref{app:ideal-dipoles}, we derive the observed Stokes parameters for the Jones matrix of ideal orthogonal dipoles. Figure~\ref{fig:induced-polarization} shows an intuitive cartoon of the induced polarization component for ideal orthogonal dipoles.

The electric field, $\bEatot$, measured by the instrument at each frequency of every spectrum is the sum of the electric fields from all sky positions, $\bEatot(\nu) = \int\bEa(\nu,\theta,\phi)\ d\Omega$. Since $\bEa(\nu,\theta,\phi)$ is zero-mean with covariance $\bSigmaa(\nu,\theta,\phi)$ and is independent at each sky position $(\theta,\phi)$, $\bEatot(\nu)$ is zero-mean with covariance $\bSigmaatot(\nu)$ where $\bSigmaatot(\nu) = \int\bSigmaa(\nu,\theta,\phi)\ d\Omega$, which implies that the Stokes parameters at each frequency are given by $P_{a,\text{tot}}(\nu) = \int P_a(\nu,\theta,\phi)\ d\Omega$. To calibrate the Stokes parameters so that antenna temperatures correspond to actual sky brightness temperatures, we consider a case where $p_s=0$ and $I_s$ is independent of angle and equal to $I_0$. In this case, the calibrated total power, $I_{a,\text{cal}}$, should be equal to $I_0$. By implementing this with a multiplicative factor, we find that $P_{a,\text{cal}}(\nu)=[2P_{a,\text{tot}}(\nu)]/[\int\Tr(\bJ^\dagger\bJ)\ d\Omega]$, i.e.
\begin{multline}
  P_{a,\text{cal}}(\nu) = \frac{\int(1-p_s)I_s\ \Tr(\bJ^\dagger\bsigma_P\bJ)\ d\Omega}{\int \Tr(\bJ^\dagger\bJ)\ d\Omega} \\ + \frac{2\int p_sI_s\bov_s^\dagger\bJ^\dagger\bsigma_P\bJ\bov_s\ d\Omega}{\int\Tr(\bJ^\dagger\bJ)\ d\Omega}. \label{eq:calibrated-antenna-stokes}
\end{multline}
Using this factor, we can also define a calibrated total electric field, $\bE_{a,\text{cal}}=\sqrt{2}\bE_{a,\text{tot}}/\sqrt{\int\Tr(\bJ^\dagger\bJ)\ d\Omega}$, and covariance matrix, $\bSigma_{a,\text{cal}}=2\bSigma_{a,\text{tot}}/\int\Tr(\bJ^\dagger\bJ)\ d\Omega$.

The Jones matrix-based formalism used here is equivalent to the commonly used Mueller matrix-based formalism. The connection between the Jones and Mueller formalisms is laid out in Appendix~\ref{app:Mueller-matrix}. It is worthwhile to note that the Mueller matrix is proportional to a product including two factors of the Jones matrix, just like both terms in Equation~\ref{eq:antenna-polarization-two-terms} have two factors of $\bJ$.

\subsection{Neglecting intrinsic polarization}

If the total intensity of the sky, $I_s$, is known, but intrinsic polarization is neglected in a prediction of the antenna Stokes parameters, then there is an unmodeled residual effect given by
\begin{subequations} \begin{align}
  \Delta_{a,\text{cal}}^{(P)} &= P_{a,\text{cal}}-\frac{\int I_s\Tr(\bJ^\dagger\bsigma_P\bJ)\ d\Omega}{\int\Tr(\bJ^\dagger\bJ)\ d\Omega}, \\
  &= \frac{\int p_sI_s\left[2\bov_s^\dagger\bJ^\dagger\bsigma_P\bJ\bov_s - \Tr(\bJ^\dagger\bsigma_P\bJ)\right]\ d\Omega}{\int \Tr(\bJ^\dagger\bJ)\ d\Omega}. \label{eq:neglecting-polarization-error}
\end{align} \end{subequations}

If an experiment has only one antenna, then $\bJ$ becomes a row vector instead of a square matrix. Defining $\bq$ as the column vector $\bJ^\dagger$, the single antenna power signal $I_a^{\text{1-ant}}$, analogous to Equation~\ref{eq:calibrated-antenna-stokes}, is given by
\begin{equation}
  I_{a,\text{cal}}^{\text{1-ant}} = \frac{\int(1-p_s)I_s|\bq|^2\ d\Omega}{\int|\bq^2|\ d\Omega} + \frac{2\int p_sI_s|\bq^\dagger\bov_s|^2\ d\Omega}{\int|\bq|^2\ d\Omega}.
\end{equation}
As in Equation~\ref{eq:antenna-polarization-two-terms}, the first (second) term represents the effects of unpolarized (polarized) sky radiation. The error term in the total power measured by a single antenna when neglecting intrinsic polarization, analogous to Equation~\ref{eq:neglecting-polarization-error}, is then given by
\begin{equation}
  \Delta_{a,\text{cal}}^{\text{1-ant}} = \frac{\int p_sI_s\cos{2\alpha}\ |\bq|^2\ d\Omega}{\int|\bq|^2\ d\Omega} \label{eq:neglecting-polarization-error-single-antenna}
\end{equation}
where $\alpha$ is defined through $|\bq^\dagger\bov_s|^2=|\bq|^2\cos^2{\alpha}$. It is clear from Equations~\ref{eq:neglecting-polarization-error}~and~\ref{eq:neglecting-polarization-error-single-antenna} that intrinsic polarization must be included in the modeling of all 21-cm signal experiments, not just those that measure Stokes parameters.

Nevertheless, due to their complexity, a complete exploration of and a process for modeling and removing effects of intrinsic polarization in global 21-cm data is left for future work.

\subsection{Stokes parameter expansion matrix}
\label{sec:expmatrix_pol}

Under the assumptions that the two polarizations have equivalent beams rotated by $90^\circ$ and are phased correctly,\footnote{These assumptions amount to $J_{\alpha x}(\theta,\phi) = J_{\alpha y}(\theta,\phi+\pi/2) = -J_{\alpha x}(\theta,\phi+\pi)$ and $\text{Im}[J^\ast_{\alpha x}(\theta,\phi)\ J_{\alpha x}(\theta,\phi-\pi/2)]=0$, where $\alpha$ is either $\theta$ or $\phi$.} isotropic intensity components like the 21-cm global signal do not induce any polarization signature. Therefore, since there are four spectra and the signal only exists in the first, the expansion matrix corresponding to data from Stokes measurements is
\begin{equation}
  \bPsi_{21,\text{Stokes}}^T = \begin{bmatrix} \bI & \bzero & \bzero & \bzero \end{bmatrix}\,,
\end{equation}
where $\bI$ and $\bzero$ are the identity and zero matrices, respectively. By providing additional data describing aspects of the foreground where the signal is known to be absent, Stokes parameter measurements provide extra leverage in the extraction of the signal.

\subsection{Averaging Stokes parameters spectra}

Normally, the Stokes parameters from $n_s$ spectra, $\{P_{a,\text{cal}}^{(1)},P_{a,\text{cal}}^{(2)},\ldots,P_{a,\text{cal}}^{(n_s)}\}$, are averaged into one, $\Paave$, through  $\Paave(\nu) = \left(\sum_{k=1}^{n_s}P_{a,\text{cal}}^{(k)}(\nu)\right)/n_s$. If the spectra are measured over a total time $\Delta t$ and the frequency resolution of each spectrum  is $\Delta\nu$, then $n_s=\Delta\nu\ \Delta t$, since $1/\Delta\nu$ is the amount of time each spectrum takes to measure.\footnote{When using drift-scan measurements, $\Delta t$ is the total integration time divided by the number of integration periods, $n_{\text{drift}}$.} Therefore,
\begin{equation}
  P_{a,\text{ave}}(\nu) = \frac{1}{\Delta\nu\ \Delta t} \sum_{k=1}^{\Delta \nu\ \Delta t}P_{a,\text{cal}}^{(k)}(\nu)\,. \label{eq:final-time-averaged-Stokes-parameters}
\end{equation}

\section{Simulations}
\label{sec:simulations}

To perform our analysis and illustrate the effects of induced polarization and 
drift-scan measurements, we need to generate two different training sets: one for the signal, described in Section~\ref{sec:signal-training-set}, and one for the beam-weighted foregrounds, described in Section~\ref{sec:beam-weighted-foreground-training-set}. While the training sets are simulated without noise, the data curves used in fits to generate the results presented in Section~\ref{sec:results} contain radiometer noise, as described in Section~\ref{sec:simulation-noise}.

\subsection{Signal training set}
\label{sec:signal-training-set}

We use the same signal training set used in Paper I of this series \citep[see Figure 2 of][]{Tauscher:18}, which was made using physical simulations from the \texttt{ares} code\footnote{https://bitbucket.org/mirochaj/ares} evaluated at frequencies of 40-120 MHz. It contains signals with troughs across this band whose depths vary between 50 and 250 mK.

If using drift-scan but not polarization measurements, the expansion matrix employed is $\bPsi_{21,\text{drift}}$, and the one employed if using polarization but not drift-scan measurements is $\bPsi_{21,\text{Stokes}}$. The full expansion matrix used for the signal when both drift-scan measurements and polarization are included is the product of the drift-scan and Stokes expansion matrices, given by
\begin{subequations} \begin{align}
    \bPsi_{21\text{,drift,Stokes}}^T &= (\bPsi_{21,\text{drift}}\bPsi_{21,\text{Stokes}})^T, \\
    &= \underbrace{\begin{bmatrix} \bI & \bzero & \bzero & \bzero & \bI & \bzero & \bzero & \bzero & \cdots \end{bmatrix}}_{n_{\text{drift}}\ \begin{bmatrix}\bI&\bzero&\bzero&\bzero\end{bmatrix}\text{'s}}.
\end{align} \end{subequations}
This reflects the fact that there are $4n_{\text{drift}}$ spectra in the data and the signal is in every fourth spectrum (i.e. the Stokes $I$ spectra). In the opposite case where neither drift-scan nor polarization measurements are used, the signal expansion matrix is simply the identity matrix.

\subsection{Beam-weighted foreground training set}
\label{sec:beam-weighted-foreground-training-set}

In principle, the beam-weighted foreground training set is created from two sources, antenna beam variations and spectral foreground maps. In this paper, as in Paper I, however, we use one foreground map,\footnote{Future work will include variations of the foreground map.} the map given by \cite{Haslam:82} scaled with a spectral index of -2.5, and many beams. The beams are defined using a Jones matrix derived from that of ideal orthogonal dipole antennas (see Appendix~\ref{app:ideal-dipoles}) modulated by an angular Gaussian whose angular scale, $\alpha$, is a function of frequency, $\nu$, allowing for beam chromaticity to be robustly included in the analysis. The full Jones matrices take the form
\begin{equation}
  \bJ = \exp{\left(-\frac{\theta^2}{4[\alpha(\nu)]^2}\right)}\  \begin{bmatrix} \cos{\theta}\cos{\phi} & -\sin{\phi} \\ \cos{\theta}\sin{\phi} & \cos{\phi} \end{bmatrix}. \label{eq:simulated-Jones-matrix}
\end{equation}
Since, as mentioned in Section~\ref{sec:antenna-polarization}, the measured Stokes parameters depend on two powers of the Jones matrix, the effective beam (i.e. the Mueller matrix; see Appendix~\ref{app:Mueller-matrix}) is proportional to $\exp{\left(-\frac{\theta^2}{2[\alpha(\nu)]^2}\right)}$.~So, the Full Width at Half Maximum (FWHM) is given by $\text{FWHM}(\nu) = \sqrt{8\ln{2}}\ \alpha(\nu)$.

\begin{figure*}[ht!]
  \centering
  \includegraphics[width=0.48\textwidth]{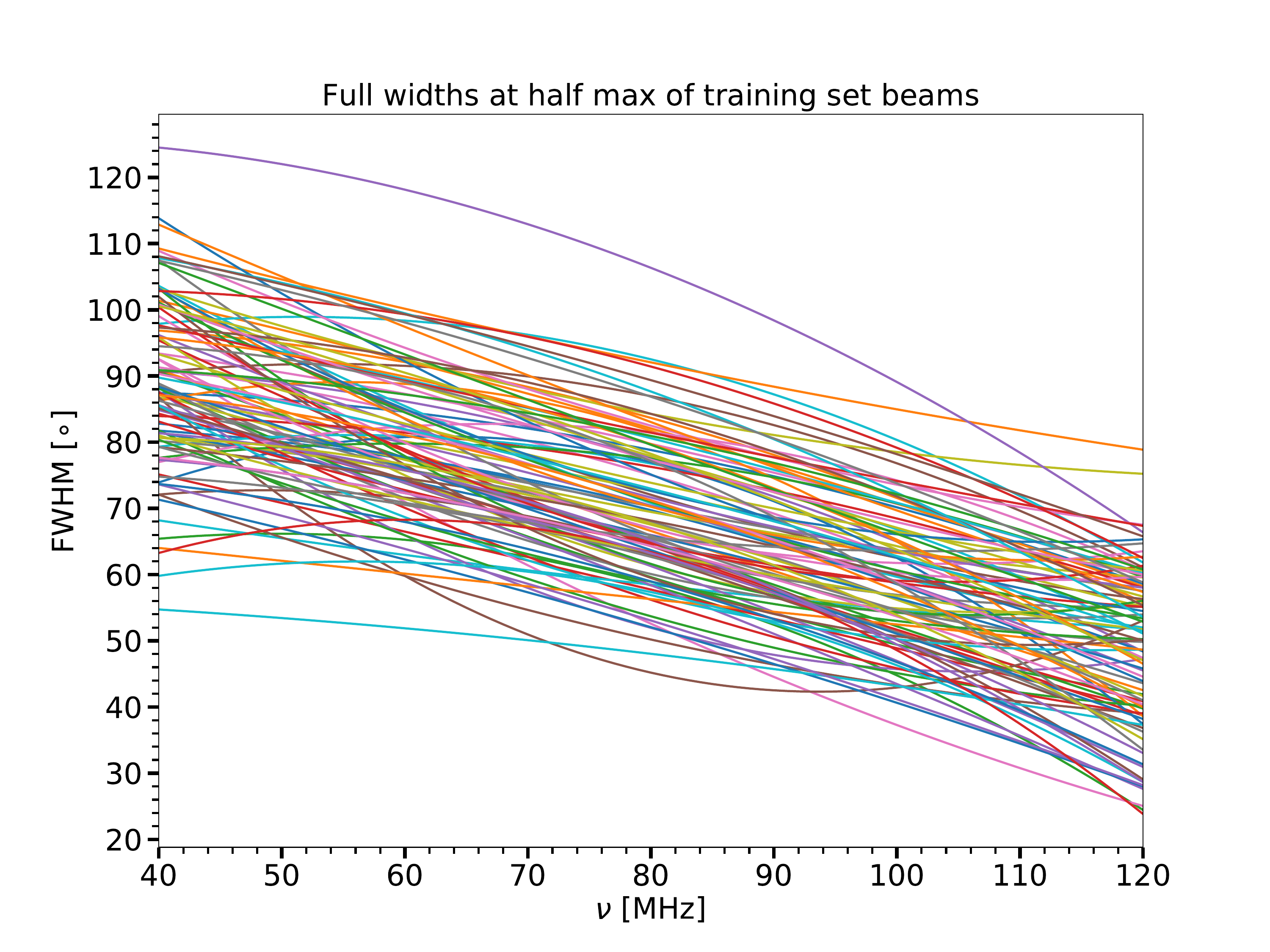}
  \includegraphics[width=0.45\textwidth]{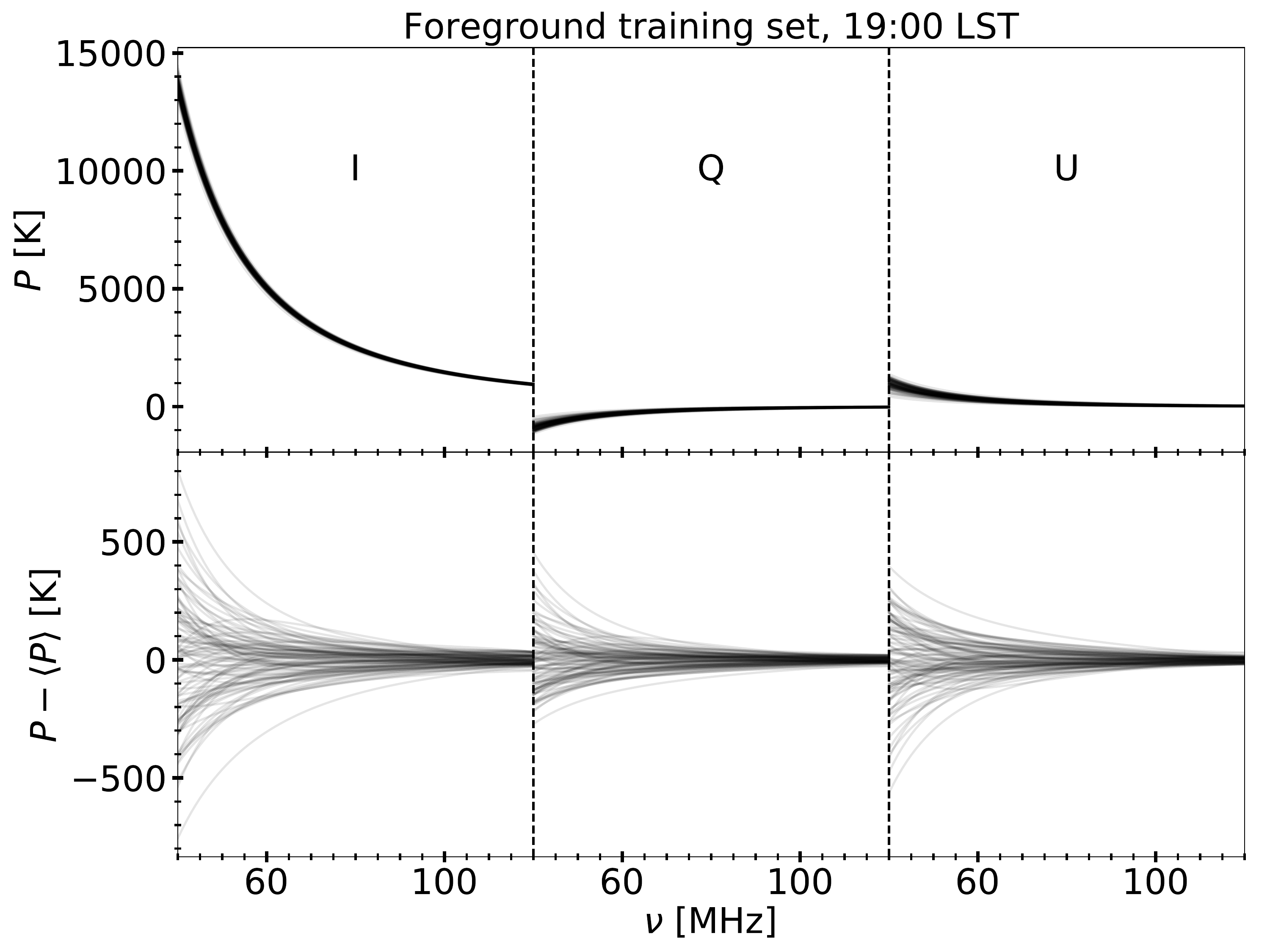}
  \caption{\textit{Left}: FWHM($\nu$) curves formed via Equation~\ref{eq:legendre-polynomials} and the distributions described in the text, Equation~\ref{eq:coefficient-distribution}, and Table~\ref{tab:legendre-coefficients}, of the Gaussian functions modulating the Mueller matrix of our simulated beam. \textit{Top right}: Training set of Stokes I, Q, and U spectra at 19:00 LST. \textit{Bottom right}: Same training set with the mean subtracted to show levels of variation.} \label{fig:foreground-training-set}
\end{figure*}

\renewcommand{\tabcolsep}{8mm}

\begin{table}[h!]
    \centering
    \caption{Parameters of Legendre coefficient distributions} \label{tab:legendre-coefficients}
    \begin{tabular}{ccc}
        \hline
        \hline
        $k$ & $\mu_k$ & $\sigma_k$ \\
        & $[^\circ]$ & $[^\circ]$ \\
        \hline
        0 & 70 & 10 \\
        1 & -20 & 5 \\
        2 & 0 & 5 \\
        \hline
        \multicolumn{3}{l}{\textbf{Note}: See Equations~\ref{eq:legendre-polynomials},~\ref{eq:legendre-polynomial-definitions},~and~\ref{eq:coefficient-distribution}.}
    \end{tabular}
\end{table}

We vary $\text{FWHM}(\nu)$ between training set elements.~For the sake of simplicity, we use $\text{FWHM}(\nu)$ curves given by quadratic polynomials in frequency. Instead of choosing the coefficients of each power of frequency independently, we utilize Legendre polynomials for easier control over the magnitude of variations, i.e.
\begin{equation}
  \text{FWHM}(\nu) = \sum_{k=0}^2 a_kL_k\left(\frac{\nu-\nu_0}{\delta\nu}\right), \label{eq:legendre-polynomials}
\end{equation}
where $\nu_0=(\nu_{\text{max}} + \nu_{\text{min}})/2$ is the average frequency, $\delta\nu=(\nu_{\text{max}}-\nu_{\text{min}})/2$ is half the width of the frequency band, and
\begin{equation}
    L_0(x) = 1,\ \ 
    L_1(x) = x,\ \ 
    L_2(x) = \frac{3x^2-1}{2}. \label{eq:legendre-polynomial-definitions}
\end{equation}
In our case, where $\nu_{\text{min}}=40$ MHz and $\nu_{\text{max}}=120$ MHz, $\nu_0=80$ MHz and $\delta\nu=40$ MHz. To seed the beam variations in our training set, we draw $a_0$, $a_1$, and $a_2$ from independent normal distributions,
\begin{equation}
  a_k \sim \mN(\mu_k, \sigma_k^2), \label{eq:coefficient-distribution}
\end{equation}
with the means and standard deviations $\mu_k$ and $\sigma_k$ given in Table~\ref{tab:legendre-coefficients}. An extra constraint is applied to exclude $\text{FWHM}(\nu)$ curves which dip below $20^\circ$ or rise above $150^\circ$ in the $40-120$ MHz band. The resulting training set of $\text{FWHM}$ curves is shown in the left panel of Figure~\ref{fig:foreground-training-set}.

We simulate observed Stokes parameters with Equation~\ref{eq:calibrated-antenna-stokes} using antenna Jones matrices of the form of Equation~\ref{eq:simulated-Jones-matrix} with the FWHM functions described above pointing at zenith from the Green Bank Observatory (GBO), the site of the Cosmic Twilight Polarimeter \citep[CTP;][]{Nhan:19}, at $38.4^\circ$ N, $79.8^\circ$ W. These simulated spectra at 19 hr LST are shown in the right panel of Figure~\ref{fig:foreground-training-set}. The full beam weighted training set includes 100 LSTs, equally spaced throughout the day.

\subsection{Simulation noise}
\label{sec:simulation-noise}

Equations~\ref{eq:Stokes-I-noise}-\ref{eq:Stokes-V-noise} of Appendix~\ref{app:Stokes-noise} describe the ideal radiometric noise level on Stokes parameter measurements from dual-antenna systems.~These equations should be used when analyzing data from a real experiment because precision in the noise level is very important for analysis accuracy.~However, since, in our case, $I_{a,\text{cal}}$ is much larger than $|Q_{a,\text{cal}}|$, $|U_{a,\text{cal}}|$, and $|V_{a,\text{cal}}|$ (see the right panel of Figure~\ref{fig:foreground-training-set}) and we are adding the noise ourselves at a known level, we simplify these equations to $\Var[P_{a,\text{ave}}]=\frac{I^2_{a,\text{ave}}}{2\Delta\nu\ \Delta t}$ for all $P\in\{I,Q,U,V\}$, meaning the standard deviation of the noise follows the simple radiometer equation with an extra factor of 2 provided by the fact that there are two independent antennas, i.e.
\begin{equation}
    \sigma_{P_{a,\text{ave}}}(\nu,t) = \frac{I_{a,\text{ave}}(\nu,t)}{\sqrt{2\Delta\nu\ \Delta t}}.
\end{equation}

For all fits in this paper, we use a total integration time of 800 hours.~When performing fits with $n_{\text{drift}}$ drift-scan measurements, we split the integration time among them equally so that $\Delta t=(800\text{ hr}) / n_{\text{drift}}$.

\section{Results}
\label{sec:results}

We perform fits to 5000 simulated data curves generated as described in Section~\ref{sec:simulations} for four different cases: full Stokes with drift-scan, full Stokes without drift-scan, Stokes I only with drift-scan, Stokes I only with no drift-scan.~When only Stokes I is used, it is assumed that the measurements are made with the same dual-antenna system as is used for full Stokes measurements so that the noise and antenna beam are the same, but Stokes Q, U, and V are simply not available. When drift-scan is used, spectra are taken from 25 foreground snapshots evenly spaced throughout the sidereal day, whereas when it is not used, all 25 of these time steps are averaged to generate the data curves to fit, which is analogous to analyzing spectra averaged over one or more full sidereal days.

To evaluate these fits, we design an RMS uncertainty to capture the bias generated through signal extraction.~To begin, we consider the $1\sigma$ RMS uncertainty defined in Equation~\ref{eq:channel-rms}. Due to overlap between signal and foreground, however, it is not guaranteed that the $1\sigma$ uncertainty interval on the signal actually contains the true signal at any particular confidence level.~To proceed, we must be able to determine the number of $\sigma$ at which the uncertainty interval of a given fit contains the input signal. This is the purpose of the signal bias statistic, $\varepsilon$, first introduced in \cite{Tauscher:18} as
\begin{equation}
  \varepsilon = \sqrt{\frac{1}{n_\nu}\sum_{i=1}^{n_\nu}\frac{[(\bgamma_{21}-\by_{21})_i]^2}{(\bDelta_{21})_{ii}}}\;, \label{eq:signal-bias-statistic}
\end{equation}
where $\by_{21}$ is the input 21-cm signal and $\bgamma_{21}$ and $\bDelta_{21}$ are given in Equations~\ref{eq:channel-mean}~and~\ref{eq:channel-covariance}.~The RMS uncertainty of the interval known to include the signal is denoted by $\text{RMS}_{21}$ and is formed by the product of Equations~\ref{eq:channel-rms}~and~\ref{eq:signal-bias-statistic},
\begin{subequations} \label{eq:confident-signal-rms}
\begin{align}
  \text{RMS}_{21} &= \varepsilon\ \text{RMS}_{21}^{1\sigma}\;, \\
  &= \frac{1}{n_\nu}\sqrt{\sum_{i=1}^{n_\nu}\sum_{j=1}^{n_\nu}\frac{(\bDelta_{21})_{jj}}{(\bDelta_{21})_{ii}}[(\bgamma_{21}-\by_{21})_i]^2}\;.
\end{align}
\end{subequations}

\begin{figure}[t!]
  \centering
  \includegraphics[width=0.47\textwidth]{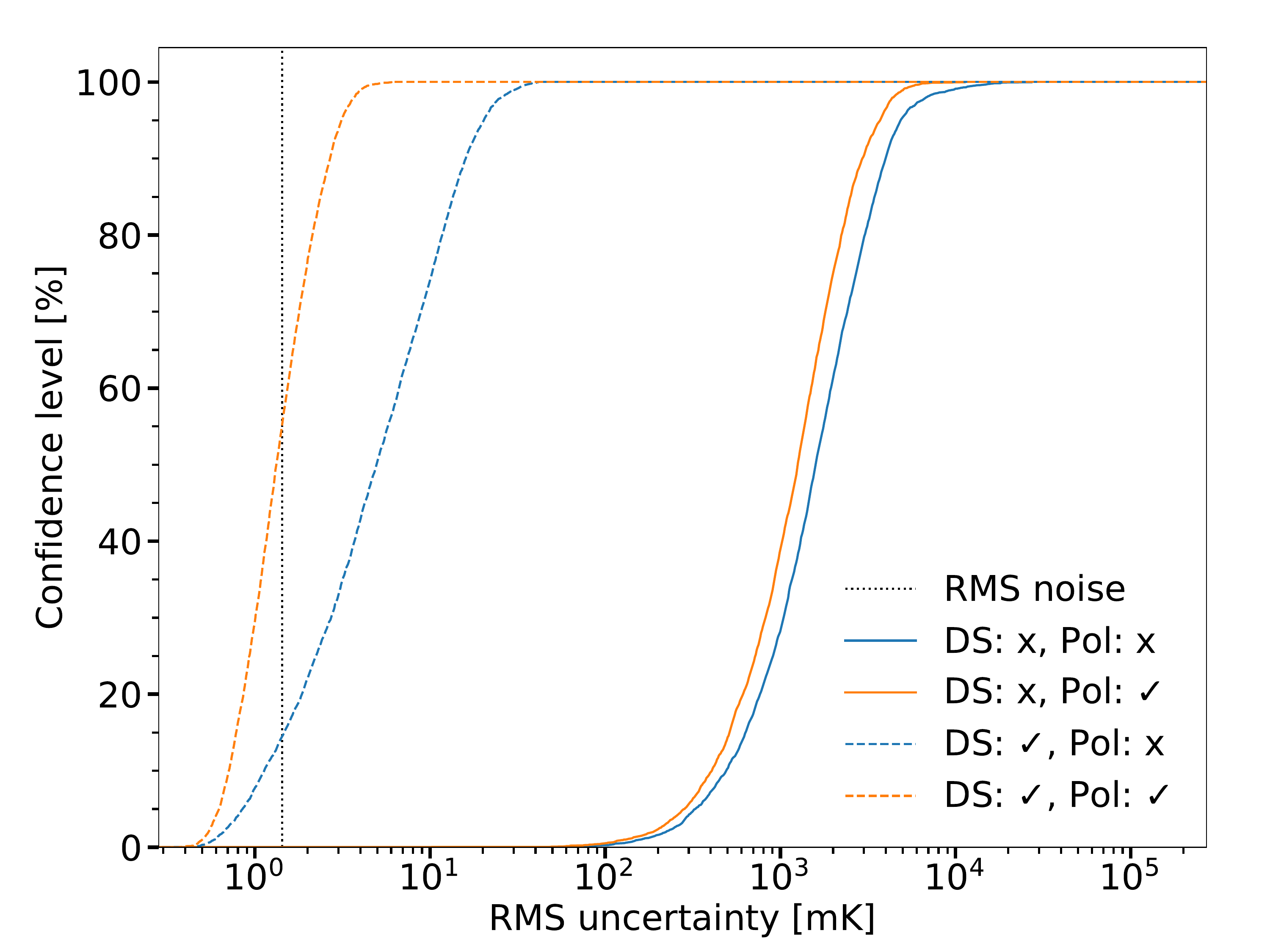}
  \caption{CDF (Equation~\ref{eq:cdf}) of the RMS uncertainty level given in Equation~\ref{eq:confident-signal-rms} from 5000 fits for each of four cases. The solid lines use a single time average for an entire sidereal day while the dashed lines break the sidereal day into 25 bins in LST, leveraging the drift-scan (DS) observation strategy advantageously. The orange lines use data from all four Stokes parameters while the blue lines use only Stokes I. The vertical, black dotted line marks the RMS noise level on the signal, i.e.~the RMS uncertainty if there was no beam-weighted foreground. This level is the same with and without drift-scan information because the same total integration time is used in both cases. The RMS uncertainty levels at 68\%, 95\%, and 99\% confidence are shown in Table~\ref{tab:confidence-levels}.} \label{fig:confidence-level-vs-rms-uncertainty}
\end{figure}

\renewcommand{\tabcolsep}{3mm}

\begin{table}[h!]
  \centering
  \caption{RMS uncertainties for different cases from Figure~\ref{fig:confidence-level-vs-rms-uncertainty}}
  \begin{tabular}{ccccc}
    \hline
    \hline
    DS & Pol & 68\% & 95\% & 99\% \\
     & & [mK] & [mK] & [mK] \\
    \hline
    %$x$ & $x$ & $2.2457\times 10^3$ & $4.8887\times 10^3$ & $1.0978\times 10^4$ \\
    %$x$ & $\checkmark$ & $1.7508\times 10^3$ & $3.6946\times 10^3$ & $5.2025\times 10^3$ \\
    %$\checkmark$ & $x$ & $8.2970\times 10^0$ & $2.0456\times 10^1$ & $3.1623\times 10^1$ \\
    %$\checkmark$ & $\checkmark$ & $1.6972\times 10^0$ & $3.1623\times 10^0$ & $4.1843\times 10^0$ \\
    $x$ & $x$ & $2.2\times 10^3$ & $4.9\times 10^3$ & $1.1\times 10^4$ \\
    $x$ & $\checkmark$ & $1.8\times 10^3$ & $3.7\times 10^3$ & $5.2\times 10^3$ \\
    $\checkmark$ & $x$ & $8.3\times 10^0$ & $2.0\times 10^1$ & $3.2\times 10^1$ \\
    $\checkmark$ & $\checkmark$ & $1.7\times 10^0$ & $3.2\times 10^0$ & $4.2\times 10^0$ \\
    \hline
    \multicolumn{5}{c}{\textbf{Note}: All values given to two significant digits.}
  \end{tabular}
  \label{tab:confidence-levels}
\end{table}

Using the values of $\text{RMS}_{21}$ for each of the 5000 fits in every case studied, we make a Cumulative Distribution Function (CDF) defined by
\begin{equation}
  \text{CDF}(x) = \Pr[\text{RMS}_{21}<x].
  \label{eq:cdf}
\end{equation}
We interpret the values of this CDF as confidence levels for future fits in which $\by_{21}$ is unknown. A CDF for each of our four cases is plotted in Figure~\ref{fig:confidence-level-vs-rms-uncertainty}. Clearly, using multiple time bins leads to more robust fits than using a single averaged spectrum and leveraging all four Stokes parameters yields better fits than using only Stokes I.

\begin{figure}[t!]
  \centering
  \includegraphics[width=0.47\textwidth]{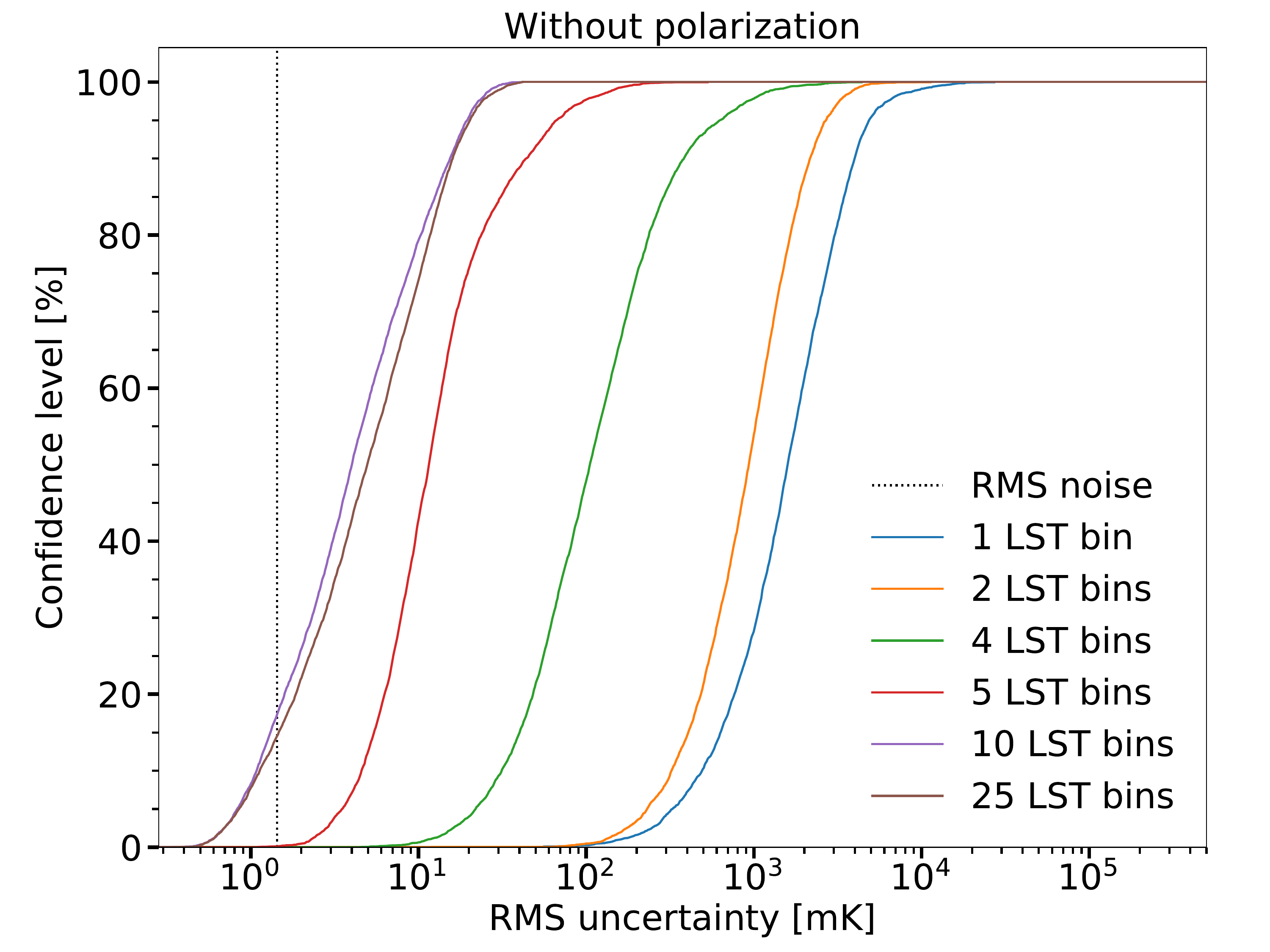}
  \includegraphics[width=0.47\textwidth]{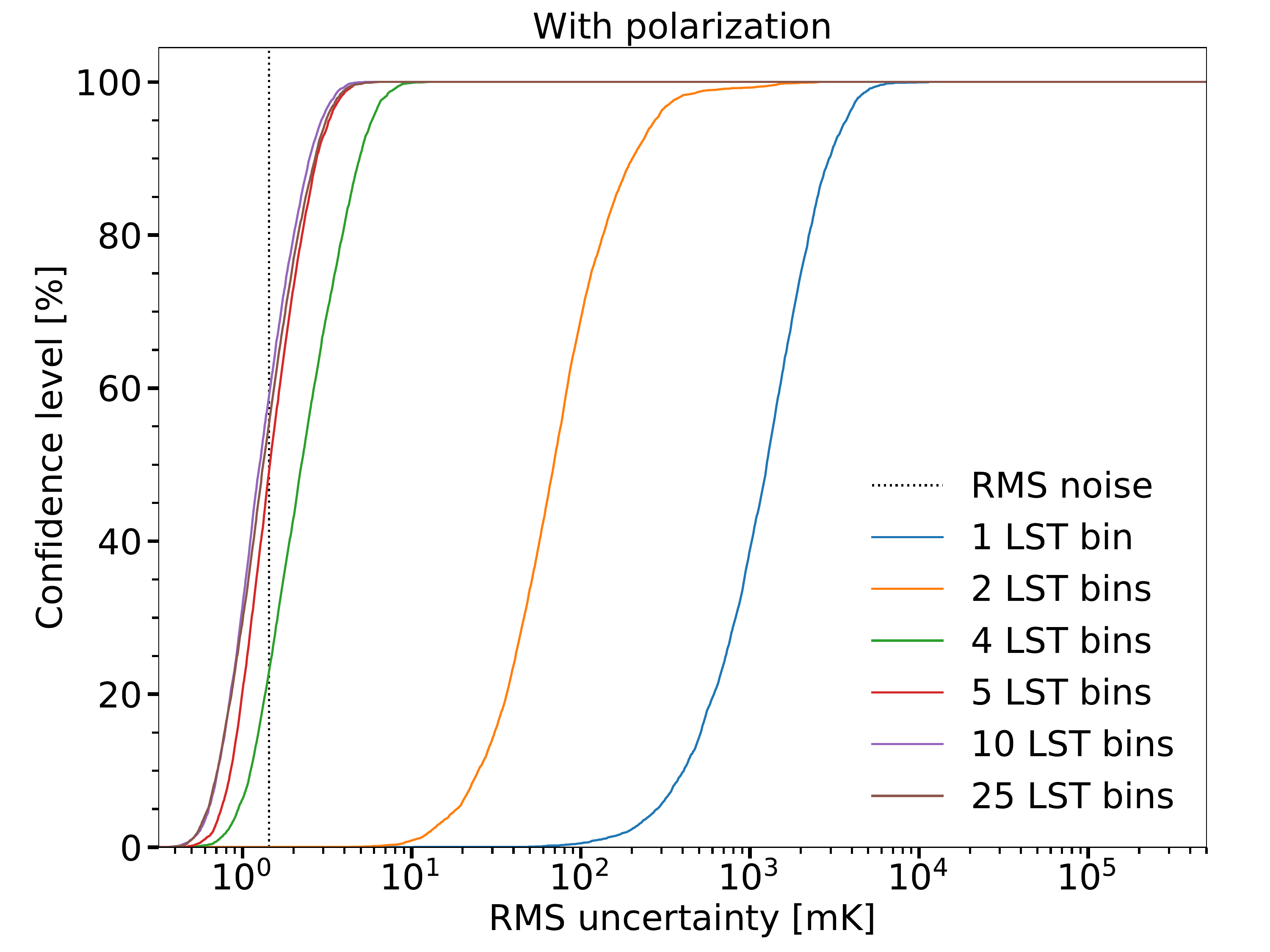}
  \caption{CDFs of the RMS uncertainty level for analyses with various numbers of LST bins, where 5000 simulations have been performed for each case. The curves in the bottom panel use data from all four Stokes parameters while the curves in the top panel use only Stokes I. The benefits of using more time bins saturate at $\sim$5-10 in both cases. This is due to the fact that the beams in the training set have FWHMs that can fit about 5 times in a 360$^\circ$ rotation (see Table~\ref{tab:legendre-coefficients} and the left panel of Figure~\ref{fig:foreground-training-set}). The vertical, black dotted lines represent the RMS noise level on the signal.} \label{fig:bias-statistic-vs-time-bins}
\end{figure}

Figure~\ref{fig:bias-statistic-vs-time-bins} shows the confidence level as a function of RMS uncertainty for various numbers of LST bins with and without polarization. In both panels, it is clear that including more time bins is necessary to achieve reasonable errors; but, they eventually saturate at around 5-10 bins due to the size of the beams used in our simulations.

\begin{figure}[t!]
  \centering
  \includegraphics[width=0.47\textwidth]{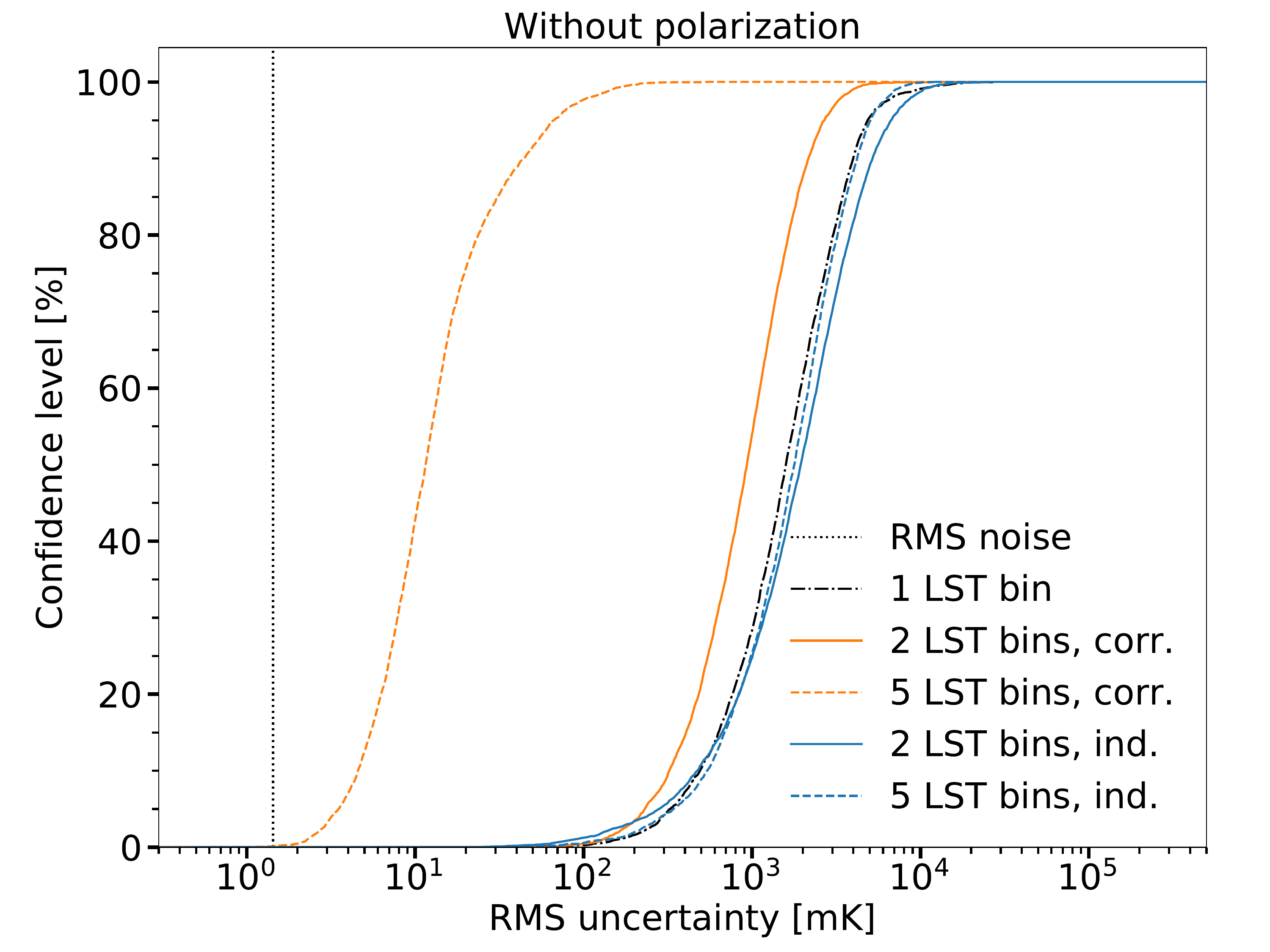}
  \includegraphics[width=0.47\textwidth]{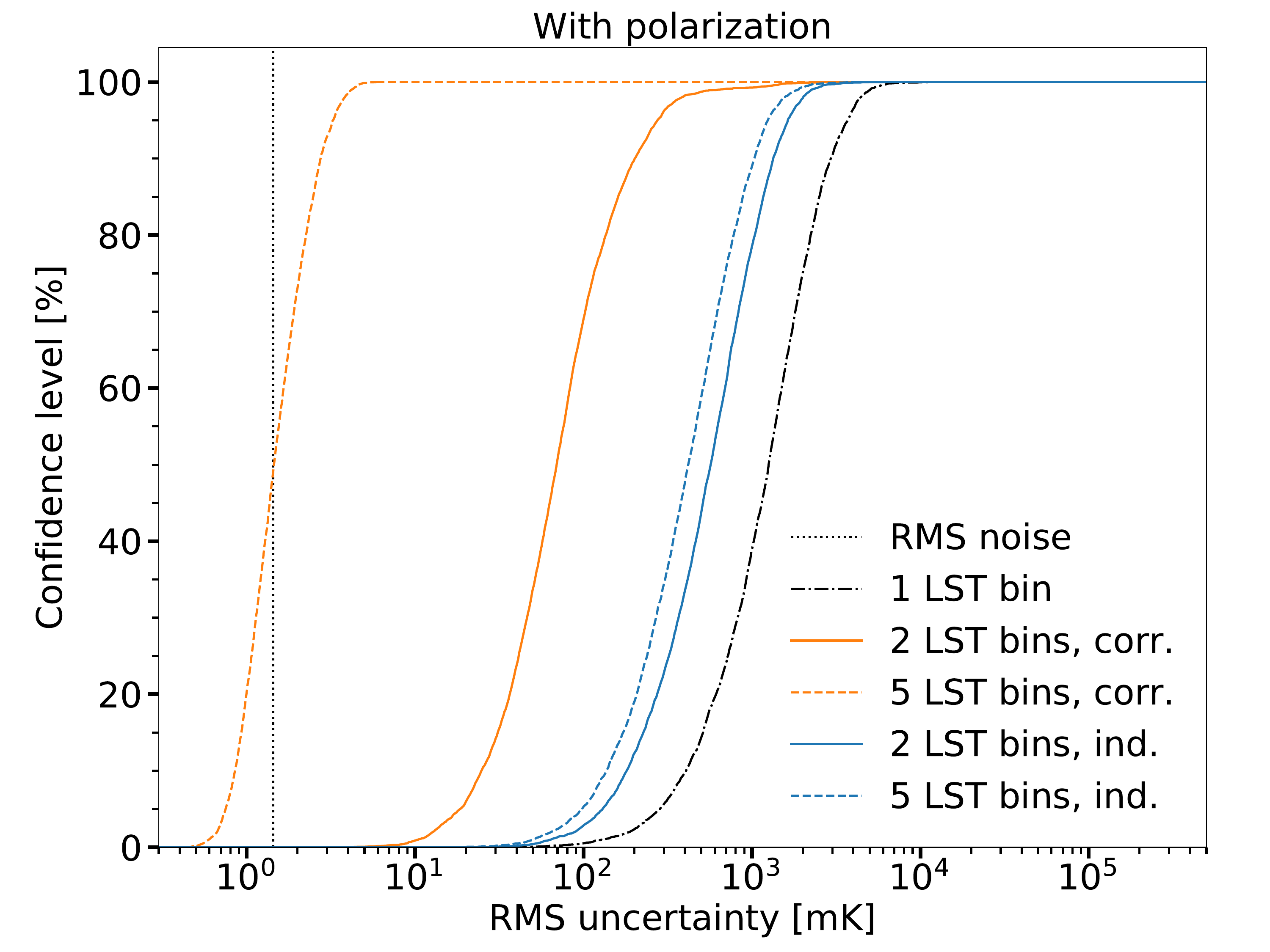}
  \caption{CDFs of the RMS uncertainty level for analyses with and without correlation assumed between time bins, where 5000 simulations have been performed for each case. The curves in the bottom panel use data from all four Stokes parameters while the curves in the top panel use only Stokes I. In Figures~\ref{fig:confidence-level-vs-rms-uncertainty}~and~\ref{fig:bias-statistic-vs-time-bins}, when multiple time bins are used, they are assumed to be correlated, as is the case here with the orange curves. The blue curves represent fits done where each spectrum has its own independent basis vectors. The black dash-dot lines represent the single time bin case and the vertical, black dotted lines represent the RMS noise level on the signal.} \label{fig:bias-statistic-vs-spectrum-correlation}
\end{figure}

So far in this paper, it has been assumed that the foreground basis vectors exist across all time bins. However, a common analysis method is to treat every spectrum as independent and model them separately, even though they are being fit simultaneously. Figure~\ref{fig:bias-statistic-vs-spectrum-correlation} shows the effects of this key difference between the two analyses. When each time bin has its own basis vectors, the benefit of using multiple time bins is severely damped. From this, it is clear that to fully benefit from fitting all spectra simultaneously, it is imperative to do so using a single matrix, with the basis vectors spanning all time bins, as opposed to using independent basis vectors in each spectrum.

\section{Conclusions}
\label{sec:conclusions}

In this third paper of the series, we defined a method for converting RMS uncertainty to a probabilistic confidence level when using the pipeline we first introduced in \cite{Tauscher:18} (Paper I). We then applied this method to different sets of simulated data representing the global 21-cm signal and foregrounds, with the purpose to test the benefits of measuring time-binned drift-scan data and Stokes parameters.

The largest impact we found was from the use of drift-scan spectra, which can be done with any global signal experiment. By using the correlations between different time bins and enforcing that the signal must be constant from spectrum to spectrum, we found that fitting multiple time bins instead of only one can decrease uncertainties from the few K level to the few mK level. It is important to note that this large benefit is not seen if using instead the traditional method where spectra are modeled independently, even if they are fit simultaneously.

Measurements of all four Stokes parameters with dual-antenna systems also proved useful in simulations to reduce uncertainties and, when in combination with the drift-scan strategy, can lead to uncertainties approaching the radiometer noise level. However, for both single and dual antenna experiments, extra care must be taken to model the effects of intrinsic sky polarization. If neither of these two independent strategies is used \citep[i.e., if analysis is done with only a single total power spectrum, such as in][]{Bowman:18}, then the uncertainties are consistently at the few K level.

\acknowledgments{We thank David Bordenave and Bang Nhan for helpful discussions on polarization measurements. We also thank Neil Bassett and Joshua Hibbard for feedback.~D.R. was supported by a NASA Postdoctoral Program Senior Fellowship at the NASA Ames Research Center, administered by the Universities Space Research Association under contract with NASA. This work is directly supported by the NASA Solar System Exploration Virtual Institute cooperative agreement 80ARC017M0006.}

%\facility{facility ID}
%\facilities{facility ID, facility ID, facility ID} 
%\software{Numpy}

\appendix

\section{Ideal dipoles}
\label{app:ideal-dipoles}

In this appendix, we consider the Jones matrix of an orthogonal pair of ideal dipoles as given by
\begin{equation}
  \bJ = \begin{bmatrix} \cos{\theta}\cos{\phi} & -\sin{\phi} \\ \cos{\theta}\sin{\phi} & \cos{\phi} \end{bmatrix}, \label{eq:intuitive-Jones-matrix}
\end{equation}
which simply encodes a geometrical projection from electric fields on the celestial sphere to electric fields on the $X$ and $Y$ antennas. This is the matrix from the original \cite{Jones:41} work (see equation 9 of that paper) generalized to account for radiation coming from directions off the zenith angle (i.e. nonzero $\theta$). Since $\bJ$ is real in this case, $\bJ^\dagger=\bJ^T$.

\subsection{Induced polarization}

The induced portion of measured polarization (first term of Equation~\ref{eq:antenna-polarization-two-terms}) is
\begin{equation}
  (P_a)_{\text{ind}} = \left(\frac{1-p_s}{2}\right)I_s\Tr(\bJ^T\bsigma_P\bJ).
\end{equation}
Computing the trace using Equation~\ref{eq:intuitive-Jones-matrix}, we find
\begin{subequations} \begin{align}
  (I_a)_{\text{ind}} &= \left(\frac{1-p_s}{2}\right)I_s(1+\cos^2{\theta}), \label{eq:total-power-induced-polarization} \\
  (Q_a+iU_a)_{\text{ind}} &= -\left(\frac{1-p_s}{2}\right)I_se^{2i\phi}\sin^2{\theta}, \label{eq:polarization-induced-polarization} \\
  (V_a)_{\text{ind}} &= 0. \label{eq:V-induced-polarization}
\end{align} \end{subequations}
Equations~\ref{eq:total-power-induced-polarization},~\ref{eq:polarization-induced-polarization},~and~\ref{eq:V-induced-polarization} with $p_s=0$ are the origin of the beam defined in Equation 9 of \cite{Tauscher:18}.

\subsection{Linear intrinsic polarization}

The intrinsic portion of measured polarization (second term of Equation~\ref{eq:antenna-polarization-two-terms}) is
\begin{equation}
  (P_a)_{\text{int}} = p_sI_s(\bJ\bov_s)^\dagger\bsigma_P(\bJ\bov_s).
\end{equation}
Using Equation~\ref{eq:intuitive-Jones-matrix} and the definition of $\bov_s$ in terms of $\psi_s$ (which applies when $V_s=0$), we find that
\begin{subequations}
\begin{align}
  \bJ\bov_s &= \begin{bmatrix} \cos{\theta}\cos{\phi} & -\sin{\phi} \\ \cos{\theta}\sin{\phi} & \cos{\phi} \end{bmatrix} \begin{bmatrix} \cos{\psi_s} \\ \sin{\psi_s} \end{bmatrix} \\
  &= \begin{bmatrix} \cos{\theta}\cos{\phi}\cos{\psi_s} - \sin{\phi}\sin{\psi_s} \\ \cos{\theta}\sin{\phi}\cos{\psi_s} + \cos{\phi}\sin{\psi_s} \end{bmatrix}.
\end{align}
\end{subequations}
This means that
\begin{subequations} \begin{align}
  (I_a)_{\text{int}} &= p_sI_s(1 - \sin^2{\theta}\cos^2{\psi_s}), \label{eq:total-power-intrinsic-polarization} \\
  (Q_a+iU_a)_{\text{int}} &= p_sI_se^{2i\phi}\ (\cos{2\psi_s} - \sin^2{\theta}\cos^2{\psi_s} + i\cos{\theta}\sin{2\psi_s}), \label{eq:polarization-intrinsic-polarization} \\
  (V_a)_{\text{int}} &= 0.
\end{align} \end{subequations}

\subsection{Combined results}

The total power seen by the antennas is given by the sum of Equations~\ref{eq:total-power-induced-polarization}~and~\ref{eq:total-power-intrinsic-polarization} while the polarization signal seen by the antenna is given by the sum of Equations~\ref{eq:polarization-induced-polarization}~and~\ref{eq:polarization-intrinsic-polarization}. After normalizing so that $I_s=I_0$ and $Q_s=U_s=V_s=0$ yield $I_{a,\text{cal}}=I_0$ (see Equation~\ref{eq:calibrated-antenna-stokes}), we find that the calibrated antenna temperatures are
\begin{subequations} \begin{align}
  I_{a,\text{cal}} &= \frac{3}{16\pi}\int I_s\ \left[(1+\cos^2{\theta}) - p_s\sin^2{\theta}\cos{2\psi_s}\right]\ d\Omega \label{eq:total-power-all-terms}, \\
  Q_{a,\text{cal}}+iU_{a,\text{cal}} &= \frac{3}{16\pi}\int I_s\ e^{2i\phi}\ \left\{ \left[p_s(1+\cos^2{\theta})\cos{2\psi_s} -\sin^2{\theta} \right] + 2ip_s\cos{\theta}\sin{2\psi_s} \right\}\ d\Omega, \\
  V_{a,\text{cal}} &= 0.
\end{align} \end{subequations}
These equations can be generalized to the case where there is intrinsic circular polarization. If it is assumed, as in the simulations of this paper, that no sky sources are intrinsically polarized ($p_s=0$), then $I_{a,\text{cal}}=\frac{3}{16\pi}\int I_s\ (1+\cos^2{\theta})\ d\Omega$ and $Q_{a,\text{cal}}+iU_{a,\text{cal}}=-\frac{3}{16\pi}\int I_s\ e^{2i\phi}\ \sin^2{\theta}\ d\Omega$.

\section{Connection to the Mueller matrix formalism}
\label{app:Mueller-matrix}

Equation~\ref{eq:Stokes-as-trace} states that, in the absence of coherent radiation, the Stokes parameters in a given basis are the trace of the product of the covariance matrix of electric fields in that basis with the Pauli matrices, $P_X=\Tr(\bSigma_X\bsigma_P)$. Since $\bsigma_P$ form a complete orthogonal basis of $2\times 2$ Hermitian matrices, subject to the inner product defined by $[\bA,\bB]=\Tr(\bA\bB)$, we can write
\begin{equation}
  \bSigma_X=\sum_{P\in\{I,Q,U,V\}}\frac{\Tr(\bSigma_X\bsigma_P)}{\Tr(\bsigma_P^2)}\bsigma_P.
\end{equation}
Since $\Tr(\bsigma_P^2)=\Tr(\bI)=2$ and $\Tr(\bSigma_X\bsigma_P)=P_X$, this means that
\begin{equation}
  \bSigma_X = \frac{1}{2}\sum_{P\in\{I,Q,U,V\}}P_X\bsigma_P. \label{eq:covariance-expanded-in-pauli-matrices}
\end{equation}
Plugging in $X=s$, multiplying on the left by $\bJ$ and on the right by $\bJ^\dagger$, and noting that $\bSigmaa=\bJ\bSigmas\bJ^\dagger$, we find
\begin{equation}
  \bSigmaa = \frac{1}{2}\sum_{P\in\{I,Q,U,V\}} P_s\  \bJ\bsigma_P\bJ^\dagger.
\end{equation}
Writing $P^\prime_a=\Tr(\bSigmaa\bsigma_{P^\prime})$ through $P^\prime_a = \sum_{P\in\{I,Q,U,V\}} P_s \mM_{P_s\rightarrow P^\prime_a}$, we can then write
\begin{equation}
  \mM_{P_s\rightarrow P^\prime_a} = \frac{1}{2}\ \Tr(\bJ\bsigma_P\bJ^\dagger\bsigma_{P^\prime}).
\end{equation}
The Mueller matrix is normalized by the integral over the $I_s\rightarrow I_a$ element, $\mM_{P_s\rightarrow P^\prime_a}^{(\text{norm})}=\mM_{P_s\rightarrow P^\prime_a}/\int\mM_{I_s\rightarrow I_a}\ d\Omega$. This normalized Mueller matrix satisfies
\begin{equation}
    \mM^{(\text{norm})}_{P_s\rightarrow P^\prime _a}(\theta,\phi,\nu) = \frac{\Tr\left\{\left[\bJ(\theta,\phi,\nu)\right]\bsigma_P\left[\bJ(\theta,\phi,\nu)\right]^\dagger\bsigma_{P^\prime}\right\}}{\int \Tr\left\{\left[\bJ(\theta,\phi,\nu)\right]^\dagger\left[\bJ(\theta,\phi,\nu)\right]\right\}\ d\Omega}.
\end{equation}
The total calibrated antenna Stokes parameters are given by
\begin{equation}
  P_{a,\text{cal}}(\nu) = \sum_{P^\prime\in\{I,Q,U,V\}} \int \mM_{P^\prime_s\rightarrow P_a}^{(\text{norm})}(\theta,\phi,\nu)\ P^\prime_s(\theta,\phi,\nu)\ d\Omega.
\end{equation}
When assuming that there are no polarized sky sources as in the simulations of this paper, the Mueller matrix effectively becomes a column vector with elements
\begin{equation}
  \mM_{I_s\rightarrow P_a}^{(\text{norm})} = \frac{\Tr\left\{\left[\bJ(\theta,\phi,\nu)\right]^\dagger\bsigma_P\left[\bJ(\theta,\phi,\nu)\right]\right\}}{\int\Tr\left\{\left[\bJ(\theta,\phi,\nu)\right]^\dagger\left[\bJ(\theta,\phi,\nu)\right]\right\}\ d\Omega}
\end{equation}
and the calibrated Stokes parameters can be written
\begin{equation}
  P_{a,\text{cal}}(\nu) = \int \mM_{I_s\rightarrow P_a}^{(\text{norm})}(\theta,\phi,\nu)\ I_s(\theta,\phi,\nu)\ d\Omega.
\end{equation}
For the orthogonal ideal dipole Jones matrix defined in Appendix~\ref{app:ideal-dipoles}, the full Mueller matrix is given by
\begin{equation}
  \mM^{(\text{norm})}(\theta,\phi) = \frac{3}{16\pi} \begin{bmatrix} 1+\cos^2{\theta} & -\sin^2{\theta} & 0 & 0 \\ -\sin^2{\theta}\cos{2\phi} & (1+\cos^2{\theta})\cos{2\phi} & -2\cos{\theta}\sin{2\phi} & 0 \\ -\sin^2{\theta}\sin{2\phi} & (1+\cos^2{\theta})\sin{2\phi} & 2\cos{\theta}\cos{2\phi} & 0 \\ 0 & 0 & 0 & 2\cos{\theta} \end{bmatrix}
\end{equation}
and the first column is the effective Mueller matrix when $p_s=0$.

\section{Noise on Stokes parameters}
\label{app:Stokes-noise}

Denoting the average of a quantity $X$ over all $n_s=\Delta\nu\ \Delta t$ spectra by $\overline{X}$, Equation~\ref{eq:final-time-averaged-Stokes-parameters} is $\Paave(\nu)=\overline{P_{a,\text{cal}}(\nu)}$. The squared noise level on the averaged, measured Stokes parameters is given by
\begin{subequations} \begin{align}
  \Var[\Paave(\nu)] &= \Var\left[\frac{1}{\Delta\nu\ \Delta t}\sum_{k=1}^{\Delta\nu\ \Delta t}P_{a,\text{cal}}^{(k)}(\nu)\right], \\
  &= \frac{\overline{\Var[P_{a,\text{cal}}(\nu)]}}{\Delta\nu\ \Delta t}, \\
  &= \frac{\overline{\langle P^2_{a,\text{cal}}(\nu) \rangle - \left(\left\langle P_{a,\text{cal}}\right\rangle\right)^2}}{\Delta\nu\ \Delta t}, \\
  &=\frac{\overline{\left\langle\left(\bE_{a,\text{cal}}^\dagger\bsigma_P\bE_{a,\text{cal}}\right)^2\right\rangle - \left(\left\langle\bE^\dagger_{a,\text{cal}}\bsigma_P\bE_{a,\text{cal}}\right\rangle\right)^2}}{\Delta\nu\ \Delta t}, \label{eq:variance-of-Paave-as-expectation-values}
\end{align} \end{subequations}
where $\langle\ldots\rangle$ represents the expectation value and $\Var[\ldots]$ represents the variance. Because the electric field $\bE_{a,\text{cal}}^{(k)}$ follows a complex normal distribution with zero mean and covariance $\bSigma_{a,\text{cal}}^{(k)}$, the expectation value of an arbitrary function of $\bE_{a,\text{cal}}^{(k)}$ is defined as
\begin{equation}
  \left\langle h\left(\bE_{a,\text{cal}}^{(k)}\right)\right\rangle=\frac{1}{\pi^2\left|\bSigma_{a,\text{cal}}^{(k)}\right|}\int h(\bx)\ e^{-\bx^\dagger\bSigma^{(k)\ -1}_{a,\text{cal}}\bx}\ d^2x, \label{eq:general-expectation-value-equation}
\end{equation}
where $\bx$ is a complex 2D vector. By performing integrals of this form, we can find that
\begin{subequations} \begin{align}
    \left\langle\bE_{a,\text{cal}}^{(k) \dagger}\bsigma_P\bE_{a,\text{cal}}^{(k)}\right\rangle &= \Tr\left[\bsigma_P\bSigma_{a,\text{cal}}^{(k)}\right], \\
    \left\langle\left(\bE^{(k) \dagger}_{a,\text{cal}}\bsigma_P\bE^{(k)}_{a,\text{cal}}\right)^2\right\rangle &= \Tr\left[\left(\bsigma_P\bSigma_{a,\text{cal}}^{(k)}\right)^2\right] + \left(\Tr\left[\bsigma_P\bSigma_{a,\text{cal}}^{(k)}\right]\right)^2.
\end{align} \end{subequations}
Plugging these expressions into Equation~\ref{eq:variance-of-Paave-as-expectation-values}, we can compute that
\begin{equation}
  \Var[\Paave(\nu)] = \frac{\overline{\Tr\left[\left(\bsigma_P\bSigma_{a,\text{cal}}\right)^2\right]}}{\Delta\nu\ \Delta t}. \label{eq:variance-before-plugging-in}
\end{equation}
Now, we write
\begin{equation}
  \bsigma_P = \begin{bmatrix} \delta_{PI} + \delta_{PQ} & \delta_{PU}-i\delta_{PV} \\ \delta_{PU}+i\delta_{PV} & \delta_{PI} - \delta_{PQ} \end{bmatrix},
\end{equation}
where $\delta_{PP^\prime}=\begin{cases}1 & P=P^\prime \\ 0 & P\ne P^\prime\end{cases}.$\footnote{Note that ${\delta_{PP^\prime}}^2=\delta_{PP^\prime}$.} This essentially encodes Equations~\ref{eq:pauli-matrices} in a single matrix. With this same definition of $\delta_{PP^\prime}$, we can write
\begin{equation}
    P_{a,\text{cal}}^{(k)} = \delta_{PI}I_{a,\text{cal}}^{(k)} + \delta_{PQ}Q_{a,\text{cal}}^{(k)} + \delta_{PU}U_{a,\text{cal}}^{(k)} + \delta_{PV}V_{a,\text{cal}}^{(k)}.
\end{equation}
Using these definitions of $\bsigma_P$ and $P_{a,\text{cal}}^{(k)}$ in Equation~\ref{eq:covariance-expanded-in-pauli-matrices}, we can write
\begin{equation}
  \bSigma_{a,\text{cal}}^{(k)} = \frac{1}{2} \begin{bmatrix} I_{a,\text{cal}}^{(k)} + Q_{a,\text{cal}}^{(k)} & U_{a,\text{cal}}^{(k)}-iV_{a,\text{cal}}^{(k)} \\ U_{a,\text{cal}}^{(k)}+iV_{a,\text{cal}}^{(k)} & I_{a,\text{cal}}^{(k)} - Q_{a,\text{cal}}^{(k)} \end{bmatrix}.
\end{equation}
Plugging these expressions into Equation~\ref{eq:variance-before-plugging-in} for Stokes I, we compute
\begin{subequations} \begin{align}
  \Var[I_{a,\text{ave}}] &= \frac{\overline{\Tr\left[\left(\bsigma_I\bSigma_{a,\text{cal}}\right)^2\right]}}{\Delta\nu\ \Delta t} \\
  &= \frac{\overline{\Tr\left[\bSigma^2_{a,\text{cal}}\right]}}{\Delta\nu\ \Delta t} \\
  &= \frac{1}{4\Delta\nu\ \Delta t} \overline{\Tr\left(\begin{bmatrix} (I_{a,\text{cal}}+Q_{a,\text{cal}})^2+U_{a,\text{cal}}^2 + V_{a,\text{cal}}^2 & \cdots \\ \cdots & U_{a,\text{cal}}^2 + V_{a,\text{cal}}^2+(I_{a,\text{cal}}-Q_{a,\text{cal}})^2 \end{bmatrix}\right)} \label{eq:sigma-squared-no-diagonal-elements} \\
  &= \frac{1}{2\Delta\nu\ \Delta t} \overline{\left[\frac{(I_{a,\text{cal}}+Q_{a,\text{cal}})^2+(I_{a,\text{cal}}-Q_{a,\text{cal}})^2}{2} + U^2_{a,\text{cal}} + V^2_{a,\text{cal}}\right]} \\
  &= \frac{\overline{I_{a,\text{cal}}^2} + \overline{Q_{a,\text{cal}}^2} + \overline{U_{a,\text{cal}}^2} + \overline{V_{a,\text{cal}}^2}}{2\ \Delta\nu\ \Delta t}, \label{eq:Stokes-I-noise}
\end{align} \end{subequations}
where the off-diagonal elements in Equation~\ref{eq:sigma-squared-no-diagonal-elements} are left out for clarity. By performing similar calculations for the other Stokes parameters, we find
\begin{subequations} \begin{align}
  \Var[Q_{a,\text{ave}}] &= \frac{\overline{I_{a,\text{cal}}^2} + \overline{Q_{a,\text{cal}}^2} - \overline{U_{a,\text{cal}}^2} - \overline{V_{a,\text{cal}}^2}}{2\ \Delta\nu\ \Delta t}, \label{eq:Stokes-Q-noise} \\
  \Var[U_{a,\text{ave}}] &= \frac{\overline{I_{a,\text{cal}}^2} - \overline{Q_{a,\text{cal}}^2} + \overline{U_{a,\text{cal}}^2} - \overline{V_{a,\text{cal}}^2}}{2\ \Delta\nu\ \Delta t}, \label{eq:Stokes-U-noise} \\
  \Var[V_{a,\text{ave}}] &= \frac{\overline{I_{a,\text{cal}}^2} - \overline{Q_{a,\text{cal}}^2} - \overline{U_{a,\text{cal}}^2} + \overline{V_{a,\text{cal}}^2}}{2\ \Delta\nu\ \Delta t}. \label{eq:Stokes-V-noise}
\end{align} \end{subequations}

\bibliographystyle{aasjournal}
\bibliography{references}

\end{document}